\documentclass[twocolumn]{aastex631}

\accepted{August 11, 2021}

\submitjournal{ApJS}

\shorttitle{Wavelength Dependence of the PRNU of the JPAS-{\it Pathfinder} Camera}
\shortauthors{Xiao et al.}

\graphicspath{{./}{figures/}}

\begin{document}

\title{The miniJPAS Survey: A Study on Wavelength Dependence of the Photon Response Non-uniformity of the JPAS-{\it Pathfinder} Camera}

\correspondingauthor{Haibo Yuan}
\email{yuanhb@bnu.edu.cn}

\author[0000-0001-8424-1079]{Kai Xiao}
\affiliation{Department of Astronomy, Beijing Normal University, Beijing, 100875, People's Republic of China}

\author[0000-0003-2471-2363]{Haibo Yuan}
\affiliation{Department of Astronomy, Beijing Normal University, Beijing, 100875, People's Republic of China}

\author{J. Varela}
\affiliation{Centro de Estudios de F{\'i}sica del Cosmos de Arag{\'o}n (CEFCA), Unidad Asociada al CSIC, Plaza San Juan 1, 44001, Teruel, Spain}

\author{Hu Zhan}
\affiliation{Key Laboratory of Space Astronomy and Technology, National Astronomical Observatories, Chinese Academy of Sciences, Beijing 100101, China}
\affiliation{Kavli Institute for Astronomy and Astrophysics, Peking University, Beijing 100871, China}

\author{Jifeng Liu}
\affiliation{National Astronomical Observatories, Chinese Academy of Sciences 20A Datun Road, Chaoyang District, Beijing, China}

\author{D. Muniesa}
\affiliation{Centro de Estudios de F{\'i}sica del Cosmos de Arag{\'o}n (CEFCA), Unidad Asociada al CSIC, Plaza San Juan 1, 44001, Teruel, Spain}

\author{A. Moreno}
\affiliation{Centro de Estudios de F{\'i}sica del Cosmos de Arag{\'o}n (CEFCA), Unidad Asociada al CSIC, Plaza San Juan 1, 44001, Teruel, Spain}

\author{J. Cenarro}
\affiliation{Centro de Estudios de F{\'i}sica del Cosmos de Arag{\'o}n (CEFCA), Unidad Asociada al CSIC, Plaza San Juan 1, 44001, Teruel, Spain}

\author{D. Crist{\'o}bal-Hornillos}
\affiliation{Centro de Estudios de F{\'i}sica del Cosmos de Arag{\'o}n (CEFCA), Plaza San Juan 1, 44001 Teruel, Spain}

\author{A. Mar{\'i}n-Franch}
\affiliation{Centro de Estudios de F{\'i}sica del Cosmos de Arag{\'o}n (CEFCA), Unidad Asociada al CSIC, Plaza San Juan 1, 44001, Teruel, Spain}

\author{M. Moles}
\affiliation{Centro de Estudios de F{\'i}sica del Cosmos de Arag{\'o}n (CEFCA), Unidad Asociada al CSIC, Plaza San Juan 1, 44001, Teruel, Spain}

\author{H. V{\'a}zquez-Rami{\'o}}
\affiliation{Centro de Estudios de F{\'i}sica del Cosmos de Arag{\'o}n (CEFCA), Unidad Asociada al CSIC, Plaza San Juan 1, 44001, Teruel, Spain}

\author{C. L{\'o}pez-Sanjuan}
\affiliation{Centro de Estudios de F{\'i}sica del Cosmos de Arag{\'o}n (CEFCA), Unidad Asociada al CSIC, Plaza San Juan 1, 44001, Teruel, Spain}

\author{J. Alcaniz}
\affiliation{Observat{\'o}rio Nacional, Minist{\'e}rio da Ciencia, Tecnologia, Inovaç{\~a}o e Comunicaç{\~o}es, 20921-400, Rio de Janeiro, RJ, Brazil}

\author{R. Dupke}
\affiliation{Observat{\'o}rio Nacional, Minist{\'e}rio da Ciencia, Tecnologia, Inovaç{\~a}o e Comunicaç{\~o}es, 20921-400, Rio de Janeiro, RJ, Brazil}

\author{C. M. de Oliveira}
\affiliation{Departamento de Astronomia, Instituto de Astronomia, Geof\'isica e Ci\^encias Atmosf\'ericas, Universidade de S\~ao Paulo, 05508-090, S\~ao Paulo, SP, Brazil}

\author{L. Sodr\'e Jr.}
\affiliation{Departamento de Astronomia, Instituto de Astronomia, Geof\'isica e Ci\^encias Atmosf\'ericas, Universidade de S\~ao Paulo, 05508-090, S\~ao Paulo, SP, Brazil}

\author{A. Ederoclite}
\affiliation{Departamento de Astronomia, Instituto de Astronomia, Geof\'isica e Ci\^encias Atmosf\'ericas, Universidade de S\~ao Paulo, 05508-090, S\~ao Paulo, SP, Brazil}

\author{R. Abramo}
\affiliation{Instituto de F\'isica, Universidade de S\~ao Paulo, 05508-090, S\~ao Paulo, SP, Brazil}

\author{N. Benitez}
\affiliation{Instituto de Astrof\'isica de Andaluc\'a - CSIC, Apdo 3004, E-18080, Granada, Spain}

\author{S. Carneiro}
\affiliation{Instituto de F\'isica, Universidade Federal da Bahia, 40210-340, Salvador, BA, Brazil}

\author{K. Taylor}
\affiliation{Instruments4, 4121 Pembury Place, La Ca\~nada Flintridge, CA 91011, USA}
 
\author{S. Bonoli}
\affiliation{Donostia International Physics Center (DIPC), Manuel Lardizabal Ibilbidea, 4, San Sebasti{\'a}n, Spain}
\affiliation{Centro de Estudios de F{\'i}sica del Cosmos de Arag{\'o}n (CEFCA), Plaza San Juan 1, 44001 Teruel, Spain}

\begin{abstract}

Understanding the origins of small-scale flats of CCDs and their wavelength-dependent variations plays an 
important role in high-precision photometric, astrometric, and shape measurements of astronomical objects.
Based on the unique flat data of 47 narrow-band filters provided by JPAS-{\it Pathfinder}, we analyze the 
variations of small-scale flats as a function of wavelength. We find moderate variations (from about $1.0\%$ at 390 nm to $0.3\%$ at 890 nm) of small-scale flats among different filters, increasing towards shorter wavelengths. Small-scale flats of two filters close in central wavelengths are strongly correlated. We then use a simple physical model to reproduce the observed variations 
to a precision of about $\pm 0.14\%$, by considering the variations of 
charge collection efficiencies, effective areas and thicknesses between CCD pixels.
We find that the wavelength-dependent variations of small-scale flats of 
the JPAS-{\it Pathfinder} camera originate from inhomogeneities of the quantum efficiency (particularly charge collection efficiency) as well as the effective area and thickness of CCD pixels. 
The former dominates the variations in short wavelengths while the latter two dominate at longer wavelengths. 
The effects on proper flat-fielding as well as on photometric/flux calibrations for 
photometric/slit-less spectroscopic surveys are discussed, particularly in blue filters/wavelengths. 
We also find that different model parameters are sensitive to flats of different wavelengths, depending on the relations between the electron absorption depth, the photon absorption length and the CCD thickness. 
In order to model the wavelength-dependent variations of small-scale flats, a small number (around ten) 
of small-scale flats with well-selected wavelengths are sufficient to reconstruct small-scale flats in other wavelengths.

\end{abstract}


\section{Introduction} \label{sec:intro}

Charge Coupled Devices (CCDs) are widely used in astronomical surveys, such as the Sloan Digital Sky Survey (\citealt{2000AJ....120.1579Y}), the ongoing Dark Energy Survey (DES; \citealt{2005ASPC..339..152W, 2016arXiv161100037D}), and the upcoming Javalambre Physics of the Accelerating Universe Astrophysical Survey (J-PAS; \citealt{2014arXiv1403.5237B}), the Multi-channel Photometric Survey Telescope (Mephisto; Er et al. 2020), the Vera Rubin Observatory (LSST; \citealt{2009arXiv0912.0201L}), the China Space Station Telescope (CSST; \citealt{2011SSPMA..41.1441Z, 2021ChinSciBull...66...11}), the Nancy Grace Roman Space Telescope (NGRST; \citealt{2021AAS...23732701M}), and the ESA Euclid telescope (\citealt{2016SPIE.9904E..0OR}). To achieve various demanding goals of the above projects, precise measurements of brightness, positions, and shapes of astronomical objects are needed, requiring proper treatment on instrumental systematics affecting photometry, astrometry, and object shape measurements.

Flat fielding is one of the most challenging steps in image processing of wide field surveys. 
It plays a key role in correcting for instrumental systematics and limiting the precision of the photometric calibration (e.g., \citealt{2006ApJ...646.1436S}).
The contributions of a flat fielding can be decomposed into two parts:  the large scale flat and the small scale flat. The former, named illumination correction sometimes, is mainly caused by the optical system of telescopes (such as the vignetting effect, field distortion) and non-uniform coating of CCD detectors. While the latter, named pixel-response non-uniformity (PRNU), is generally related to the inhomogeneities of the quantum efficiency between adjacent CCD pixels, assuming that all CCD pixels have the same size and are uniformly distributed. However, recent studies have shown that the variations of the effective area of the pixel (pixel-to-pixel size variations) play an important role in determining the PRNU of some CCD detectors (\citealt{2017PASP..129h4502B} and references therein). In this case, dividing a raw image by a flat-field image  is no longer valid for flat-fielding correction. To better understand the nature of PRNU of CCD detectors, the dependences of PRNU on wavelength are very essential.

The J-PAS survey aims to image thousands of square degrees of the northern sky with a unique set of $54$ narrow band filters, covering 3785\,\AA \ to 9100\,\AA, using a dedicated 2.55\,$m$ telescope, JST/T250, at the Javalambre Astrophysical Observatory (\citealt{2020arXiv200701910B})\footnote{\url{https://j-pas.org}}. In its commissioning phase, a pathfinder camera was firstly installed to test the telescope performance and execute the first scientific operations.

A large number of sky flat images in each filter have been obtained, providing a unique dataset to investigate the wavelength dependent PRNU of the JPAS-{\it Pathfinder} camera.

In this work, small scale flat field images for each narrow band filter are computed from co-added flat field images, and then used to study the wavelength dependent PRNU of the JPAS-{\it Pathfinder} camera. 
A simple physical model is constructed to parametrize/reproduce the variations of small scale flats as a function of wavelength, taking into account the variations of quantum efficiencies (caused by charge collection efficiencies), pixel sizes and depths.

The paper is organized as follows. In Sections \ref{sec:data} and \ref{sec:model}, we introduce the data and model used in this work.
The results are presented in Section \ref{sec:results} and discussed in Section \ref{sec:discussion}. A summary is given in Section \ref{sec:conclusions}.

\section{Data} \label{sec:data}

The JPAS-{\it Pathfinder} camera, located at the center of the JST/T250 field-of-view, is equipped with a single large, $9216\times9232$ CCD290-99 detector from Teledyne e2V.
The detector has 16 outputs for fast read out. It has an imaging area of 92.16\,$mm$ $\times$ 92.32\,$mm$, corresponding to a $0.27$ deg$^2$ field-of-view. The pixel size is 10\,$\mu m$ and depth is 40\,$\mu m$.  
A broadband anti-reflective coating is adopted to optimize performance from 
380\,nm to 850\,nm.
Other technical parameters of the detector can be found in Table 2 of \citet{2020arXiv200701910B}.

The novel and unique aspect of J-PAS lies in its filter system: $54$ narrow band filters ranging from 3780\,$\text{\AA}$ to 9100\,$\text{\AA}$, complemented with two broader filters in the blue and red wavelength regions. The narrow band filters have a FWHM of 145\,$\text{\AA}$ and are spaced by about 100\,$\text{\AA}$ (except for the filter J$0378$), thus covering the entire optical range.
The blocking of the filters is better than OD5 (transmission $< 10^{-5}$) in the range 250 to 1050 nm  (\citealt{2018JATIS...4a5002B}),
therefore, the photometric leakage in the filter bandpasses is not important.
The J-PAS filters have also been designed to minimize internal reflections. The intensity of the parasitic light shall at least six orders of magnitude smaller than the incident light. 
These low internal reflections are not measurable in real images (\citealt{2018JATIS...4a5002B}).

Due to the low sky background of narrow band filters, the CCD scientific images are read out in a $2\times2$ binning mode to reduce 
the readout noise, as were the flat images. Therefore, each pixel in the flat images corresponds to four physical pixels.

\begin{table}[ht!]
\centering
\caption{Numbers of exposure times for different filters}
\begin{tabular}{cc|cc|cc}
\hline
\hline
Filter  & Number  & Filter   & Number & Filter   & Number \\ \hline
J0390    &     17  &  J0590    &     36 &    J0800$^{\it a}$    & 5  \\
J0400    &     52  &  J0600    &     17 &    J0810$^{\it a}$    & 7  \\
J0410    &     36  &  J0610    &     34 &    J0820$^{\it a}$    & 10 \\
J0420    &     30  &  J0620    &     36 &    J0830                 & 17 \\
J0430    &     29  &  J0630    &     37 &    J0840                 & 17 \\
J0440    &     65  &  J0640    &     29 &    J0850                 & 17 \\
J0450    &     50  &  J0650    &     83 &    J0860                 & 42 \\
J0460    &     17  &  J0660    &     32 &    J0870                 & 55 \\
J0470    &     46  &  J0670    &     17 &    J0880                 & 55 \\
J0480    &     35  &  J0680    &     35 &    J0890                 & 55 \\
J0490    &     32  &  J0700    &     60 &    J0900$^{\it b}$    & 57 \\
J0500    &     25  &  J0710    &     29 &    J0910$^{\it b}$    & 56 \\
J0510    &     62  &  J0720    &     74 &    J1007$^{\it b}$    & 54 \\
J0520    &     51  &  J0730    &     37 &    \\
J0530    &     17  &  J0740    &     17 &    \\
J0540    &     46  &  J0750    &     54 &    \\
J0550    &     36  &  J0760    &     36 &    \\
J0560    &     31  &  J0770    &     60 &    \\
J0570    &     34  &  J0780    &     29 &    \\
J0580    &     79  &  J0790    &     71 &    \\\hline 

\end{tabular}
\\
$^{a}$Not used due to small numbers of exposure times.\\
$^{b}$Not used due to fringing patterns. 

\label{exposure number}
\end{table}

Twilight flats are used for the flat fielding of mini-J-PAS observations.
Several to tens of sky flats are usually obtained for each filter, and then co-added to obtain the master flat after bias subtraction. 

The typical SNRs are about 1000 per pixel. 
For each master flat, its large-scale flat is estimated by a running mean filter smoothing,
with window size of 50 $\times$ 50 pixels.
Its small-scale flat is obtained by dividing the original flats by the large-scale one. 

The numbers of exposures for different filters are listed  in Table \ref{exposure number}. The flats for three filters (J0800, J0810, and J0820) are excluded in the current work due to low signal-to-noise ratios. The flats of another three filters (J0900, J0910,
and J1007) are also excluded,  due to strong fringing patterns in the flat images. 
J0360 flat is excluded due to its wider wavelength coverage. Above, in total the flats of 47 narrow-band filters are used for the calculation of small-scale variation.

\section{Model} \label{sec:model}

\begin{table}[ht!]
\centering
\caption{Photon absorption length at different wavelengths of Si at -$100^{\circ}$C (\citealt{1979SSEle..22..793R}, \citealt{1995Optical})}
\begin{tabular}{cc|cc|cc}
\hline
\hline
$\lambda$(nm) & L($\mu$m) & $\lambda$(nm) & L($\mu$m) & $\lambda$(nm) & L($\mu$m) \\ \hline
250 & 0.006409 & 460 & 0.8082 & 670 & 6.916 \\
260 & 0.005938 & 470 & 0.9823 & 680 & 7.479 \\
270 & 0.005326 & 480 & 1.125  & 700 & 8.776 \\
280 & 0.004886 & 490 & 1.277  & 710 & 9.466 \\
290 & 0.005116 & 500 & 1.424  & 720 & 10.14 \\
300 & 0.006589 & 510 & 1.621  & 730 & 10.99 \\
310 & 0.007883 & 520 & 1.790  & 740 & 11.99 \\
320 & 0.008840 & 530 & 2.011  & 750 & 13.17 \\
330 & 0.009656 & 540 & 2.245  & 760 & 14.48 \\
340 & 0.01037  & 550 & 2.482  & 770 & 15.77 \\
350 & 0.01091  & 560 & 2.751  & 780 & 17.30 \\
360 & 0.01124  & 570 & 2.996  & 790 & 18.97 \\
370 & 0.01685  & 580 & 3.27   & 830 & 28.23 \\
380 & 0.04285  & 590 & 3.569  & 840 & 31.24 \\
390 & 0.09592  & 600 & 3.882  & 850 & 34.93 \\
400 & 0.1713   & 610 & 4.231  & 860 & 39.43 \\
410 & 0.2438   & 620 & 4.594  & 870 & 44.44 \\
420 & 0.3314   & 630 & 4.961  & 880 & 50.90 \\
430 & 0.4263   & 640 & 5.355  & 890 & 57.82 \\
440 & 0.5414   & 650 & 5.814 \\
450 & 0.6642   & 660 & 6.356 \\ \hline
\end{tabular}
\label{absorption_length}
\end{table}

\begin{figure}[h]
\centering
   \includegraphics[width=7cm]{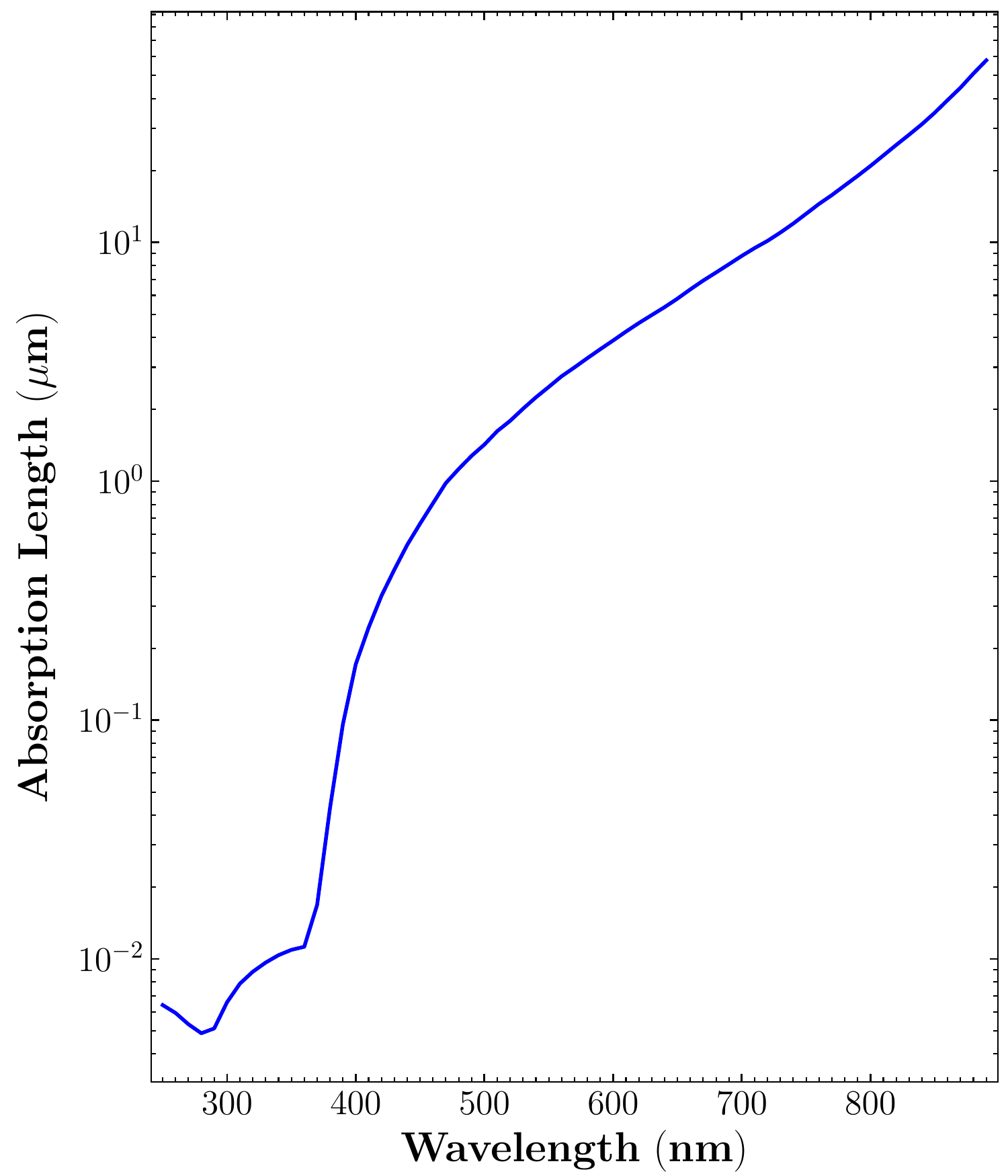}
\caption{{\small Dependence on the wavelength of the absorption length in Si at -$100^{\circ}$C is shown (\citealt{1979SSEle..22..793R}, \citealt{1995Optical}).
}}
  \label{Fig:absorption_Length}
\end{figure}

The analysis of PRNU of the CCDs can be analyzed from the perspective of the photovoltaic conversion process. Ignoring the charge loss during the charge transfer process, the electrical signal generated on a CCD pixel by monochromatic light can be expressed as 
\begin{equation}
  I_{\rm i} = \frac{1}{gain}\frac{\eta_{\rm i}A_{\rm i}}{h\nu}\int_{t}Edt,
\label{eq:Model_formula1}
\end{equation}
where $\eta_{\rm i}$ and $A_{\rm i}$ are the quantum efficiency and the effective area of the $i$\,th pixel, respectively; $h$, $\nu$, $E$, and $t$ 
are the Planck constant, frequency of radiated photons, irradiance of the CCD surface, and exposure time, respectively. The $gain$ of a CCD is set by the output electronics and determines how the amount of charges collected in each pixel will be assigned to a digital number in the output image \citep[e.g.,][]{2006hca..book.....H}. 

The quantum efficiency of a CCD pixel can be expressed as
\begin{equation}
  \eta_{\rm i} = \alpha \rm{CCE}_{\rm i} (1-R)(1-\emph{e}^{-H_{\rm i}/L}),
\label{eq:Model_formula2}
\end{equation}
where $\alpha$ is the quantum yield, which is related to photon energy; $\rm{CCE}_{\rm i}$ is the charge collection efficiency, $R$ is the reflectivity of the CCD, $H_{\rm i}$ is the thickness of silicon photosensitive layer, and $L$ is the photon absorption length (\citealt{1985SPIE..570....7J,2007ptd..book.....J}).

\citet{2014SPIE.9154E..16V} found that small-scale flats show much larger variations in shorter wavelengths 
for a BACKSIDE-ILLUMINATED CCD. Its PRNU is dominated by the residual step pattern of imperfect ion implantation and laser annealing. The pattern is strongest at the shortest wavelengths, for which the photon absorption depth is shallowest. Therefore, it is likely that the PRNU in the near ultraviolet and blue is dominated by non-uniform trapping or recombination of photon-generated electrons in the very top layer of the CCD back surface (i.e., those "partial events" defined in \citealt{1985SPIE..570....7J}), which may be thought of as an effective absorption. Based on this assumption, Chen \& Zhan (private communication) and Du et al. (in prep) proposed four models of $\rm{CCE}_{\rm i}$ to describe the probability of the electrons being "absorbed" as a function of depth in the thin layer. 
The models were tested using flat fields in the lab in wavelengths from 360\,$nm$ to 625\,$nm$.
It was found that the
model assuming the probability of an electron to be absorbed following an exponential decay matches with the data best.
Therefore, we adopt the same model in this work.  
According to the exponential decay model, the charge collection efficiency of the i-th pixel can be written as 
\begin{equation}
  \rm{CCE}_{\rm i} = 1-\frac{P_{\rm i}d_{\rm i}}{L+d_{\rm i}}\frac{1-\emph{e}^{-(L+d_{\rm i})H_{\rm i}/Ld_{\rm i}}}{1-\emph{e}^{-H_{\rm i}/L}},
\label{eq:Model_formula3}
\end{equation}
where $d$ is the decay scale length and $P$ the probability of being absorbed per unit depth at the CCD surface.

\begin{figure}
   \resizebox{\hsize}{!}{\includegraphics{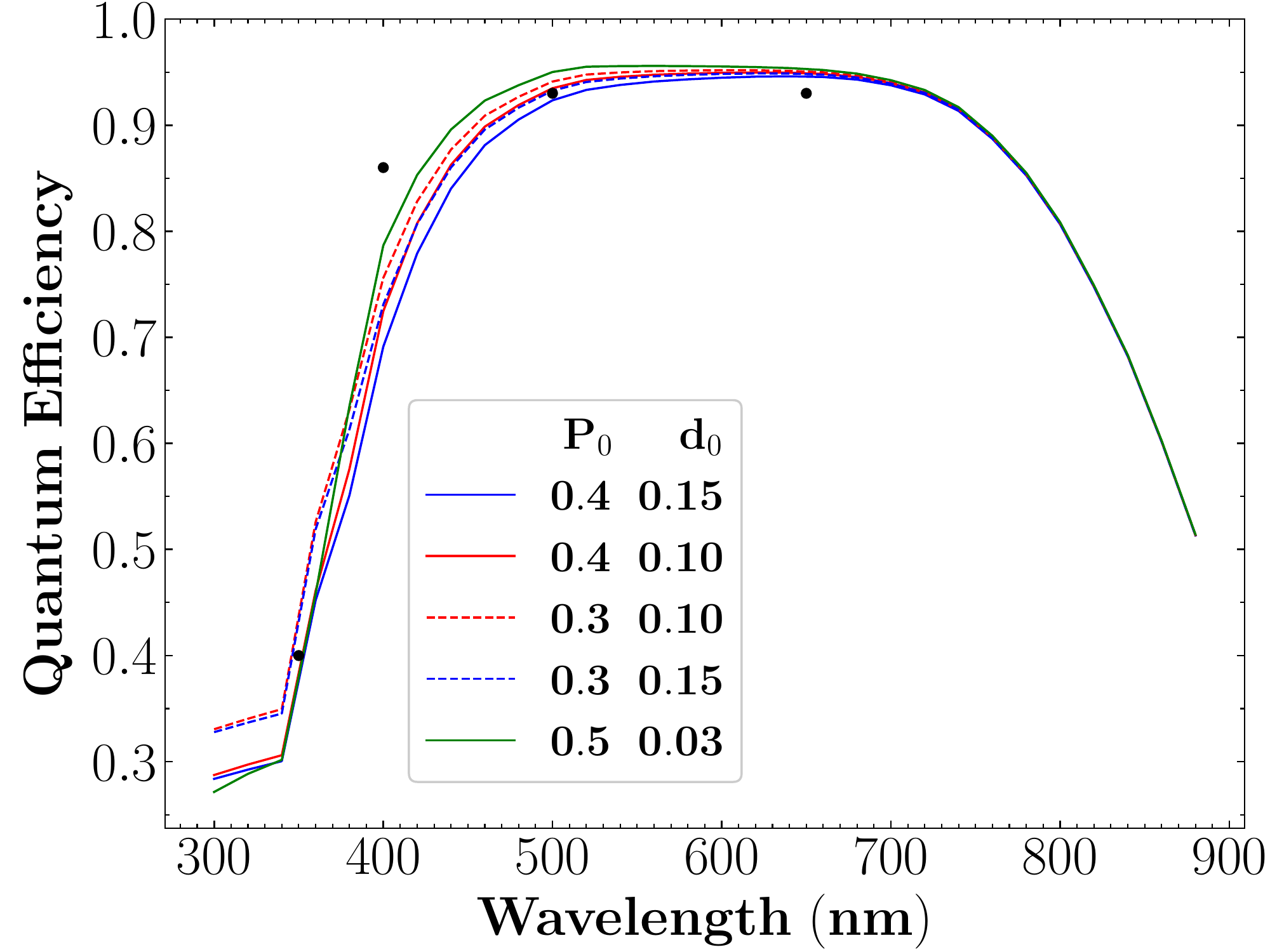}}
   \caption{{\small Quantum efficiency as a function of wavelength for different [P, d] combinations, assuming typical 
   anti-reflectivity coating efficiency factors. The black dots are quantum efficiencies of the mini-JPAS camera (\citealt[Table~2]{2020arXiv200701910B}). 
} }
  \label{Fig:QE_0401}
\end{figure}  

Equations (\ref{eq:Model_formula1})-(\ref{eq:Model_formula3}) are in dimensional form. If we neglect the CCD reflectivity non-uniformity, which is probably not important on the
    50 $\times$ 50 pixel scales,  and we adopt the mean electrical signal
    of each pixel $\bar I=\frac{1}{N}\sum\limits_{\rm{i}=0}^{N-1}I_{\rm i}$ as normalizing constant,
    we can obtain the dimensionless equation from Equations (\ref{eq:Model_formula1})-(\ref{eq:Model_formula3}) as
\begin{equation}
  \frac{I_{\rm i}}{\bar I} = \frac{A_{\rm i}\rm{CCE}_{\rm i}(1-\emph{e}^{-H_{\rm i}/L})}{\frac{1}{N}\sum\limits_{\rm{i}=0}^{N-1}A_{\rm i}\rm{CCE}_{\rm i}(1-\emph{e}^{-H_{\rm i}/L})}.
\label{eq:Model_formula4}
\end{equation}

\begin{figure*}
\begin{minipage}[t]{0.5\linewidth}
\centering
\resizebox{\hsize}{!}{\includegraphics{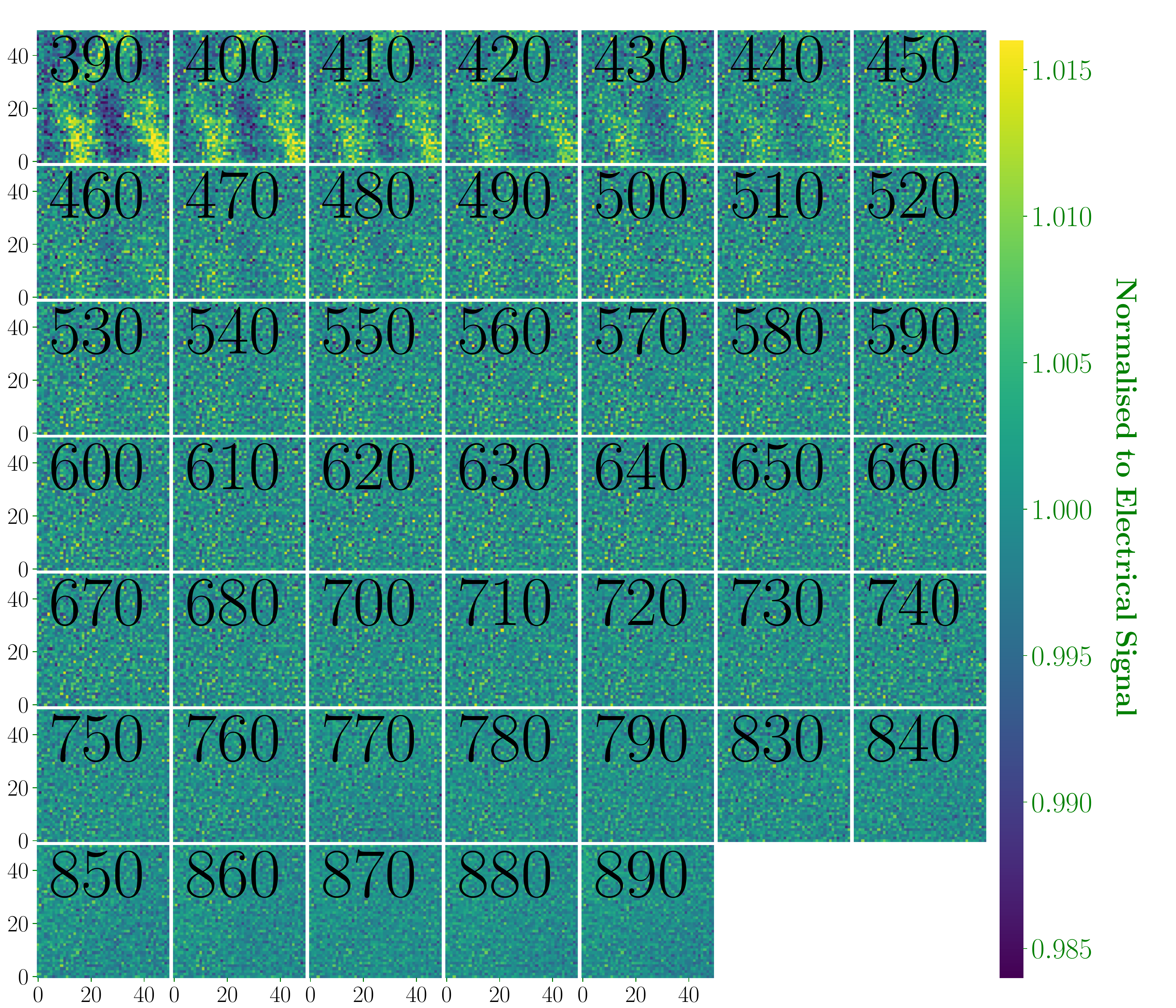}}
\end{minipage}
\begin{minipage}[t]{0.5\linewidth}
\centering
\resizebox{\hsize}{!}{\includegraphics{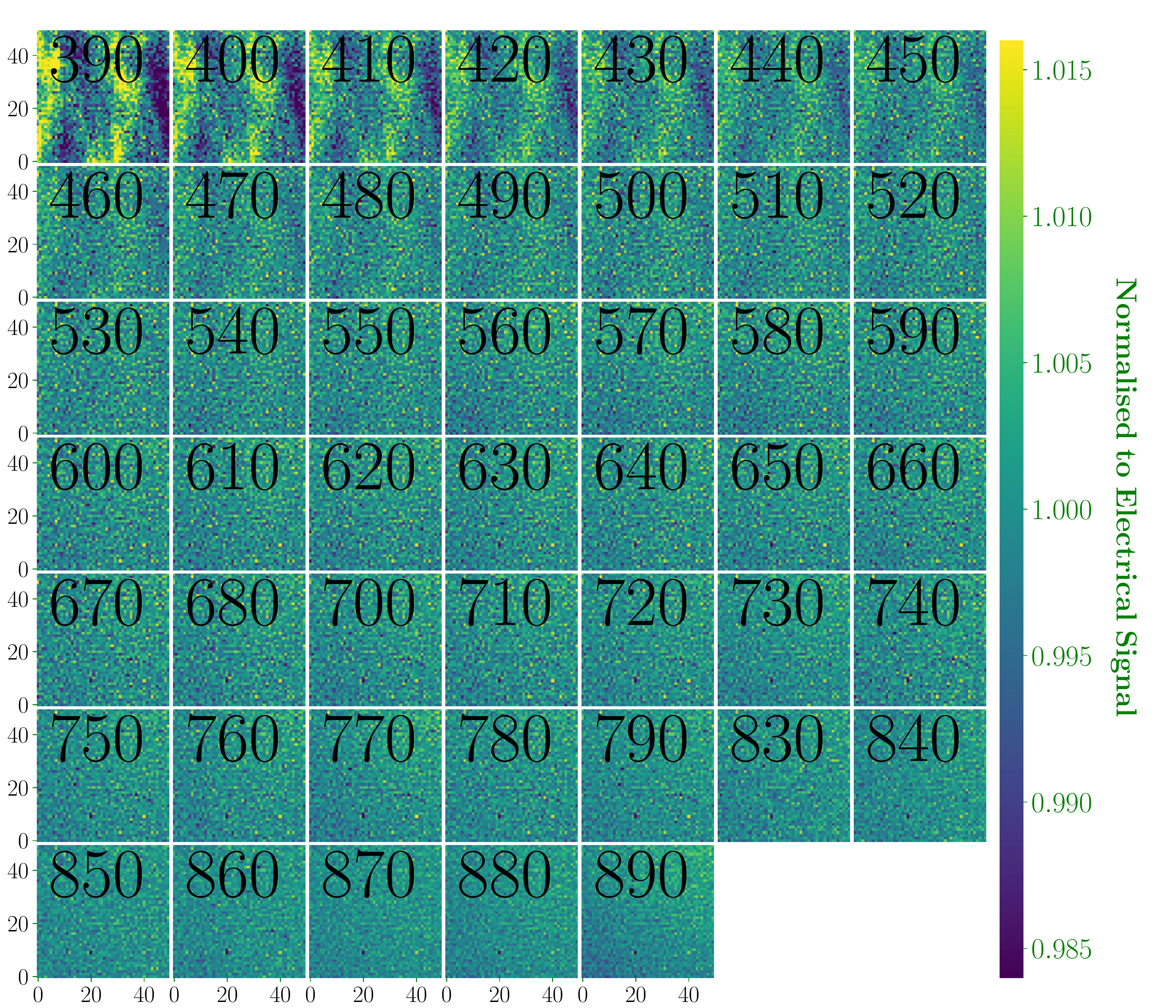}}
\end{minipage}
\caption{\small Small-scale flats at 47 wavelengths for selected areas are composed of 50 $\times$ 50 pixels. The left panel is for center region (X from 1786 pixel to 1836 pixel and Y from 2132 pixel to 2182 pixel) and the right panel is for the lower-left corner (X from 245 pixel to 295 pixel and Y from 245 pixel to 295 pixel). }
\label{Fig:obs_2500}
\end{figure*}

For a given wavelength, the photon absorption length in Si depends only on temperature.
    A theoretical calculation of temperature-dependent photon absorption lengths was given in \citet{1979SSEle..22..793R}. 
    The actual measured values of the photon absorption lengths in Si at 300K can be found in \citet{1995Optical}.
    However, at most wavelengths for this work, the discrepancies between the calculated and measured photon absorption lengths at 300K are larger (by about $30\%$). 
    
	Therefore, assuming that the differences do not depend on temperature, the photon absorption lengths used in the work are obtained as follows. We first calculate the theoretical ratios between the photon absorption lengths at 173\,$K$ (working temperature of the mini-JPAS camera) 
    and 300\,$K$, then multiply by the measured photon absorption lengths at 300\,$K$. The adopted photon absorption lengths of Si at 173\,$K$ as a function of wavelength are given in Table \ref{absorption_length} and plotted in 
    Figure\,\ref{Fig:absorption_Length}. It can be seen that the absorption length does not vary substantially for wavelength between 250\,$nm$ to 370\,$nm$, but increases rapidly to 420\,$nm$, and then more slowly to 890\,$nm$.
 
For a given wavelength, there are only four free parameters ($P_{\rm i}$, $d_{\rm i}$, $H_{\rm i}$, and $A_{\rm i}$) for a given pixel in Equation (\ref{eq:Model_formula4}). 
Therefore, the values of these parameters are possibly well constrained by a small number ($\ge4$) of small-scale flats. This provides the possibility that a small number ($\ge4$) of small-scale flats with well-selected wavelengths are sufficient to reconstruct small-scale flats in other wavelengths. Note that $A_{\rm i}$ is the effective area, and is used to account for wavelength-independent variations of small-scale flats. The variations of $A_{\rm i}$ do not necessarily mean the variations in physical size.  

\begin{figure}
   \resizebox{\hsize}{!}{\includegraphics{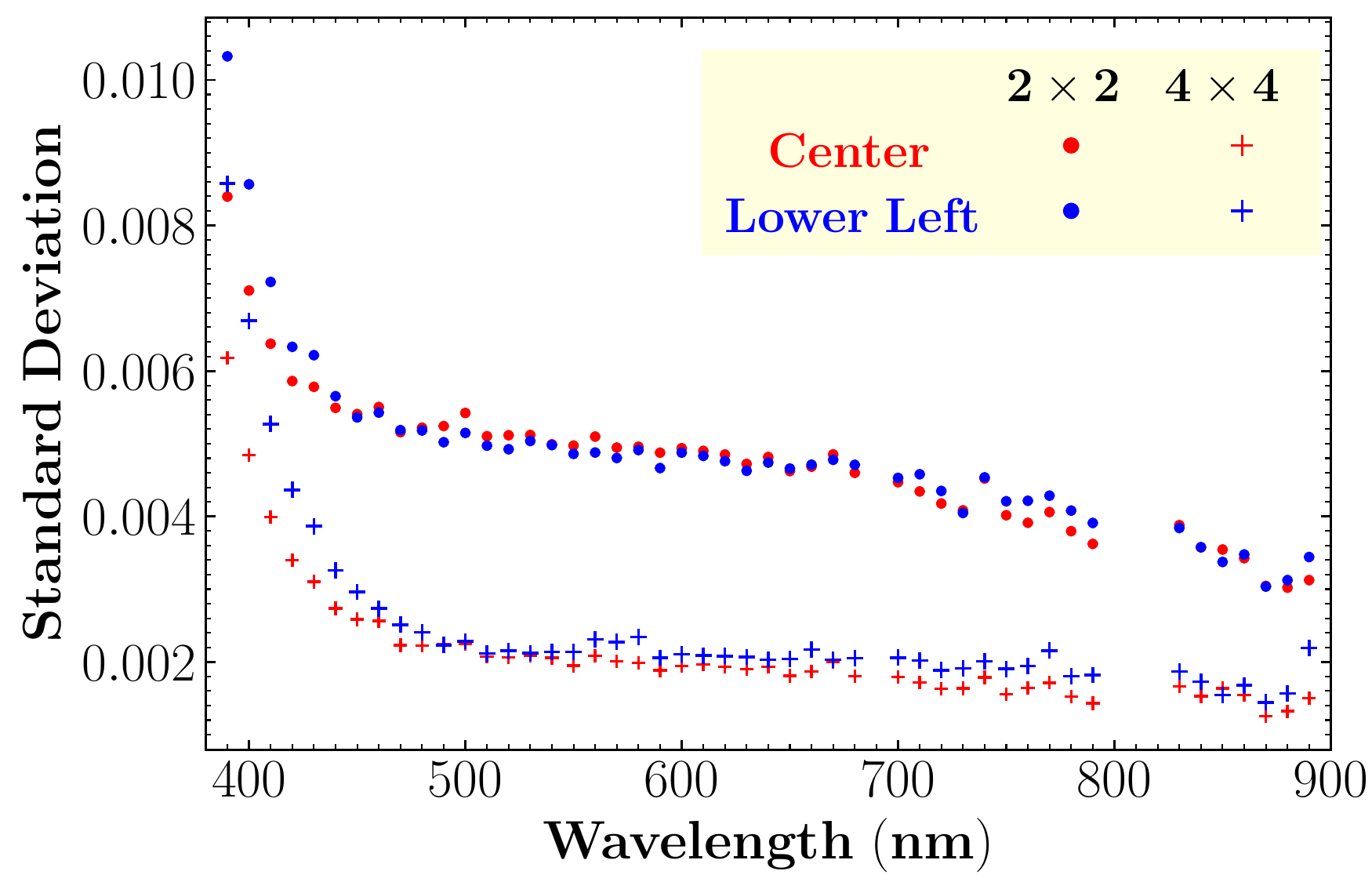}}
   \caption{{\small Standard deviations of small-scale flats as a function of wavelength. The red and blue symbols 
   are for the center and lower-left corner regions, respectively. The dots denote results after a $2\times2$ binning during the readout process. The crosses denote results in an additional $2\times2$ binning after readout, i.e., $4\times4$ binning in physical pixel. 
} }
  \label{Fig:sigma_flat}
\end{figure}

\begin{figure*}
\begin{minipage}[t]{0.5\linewidth}
\centering
\resizebox{\hsize}{!}{\includegraphics{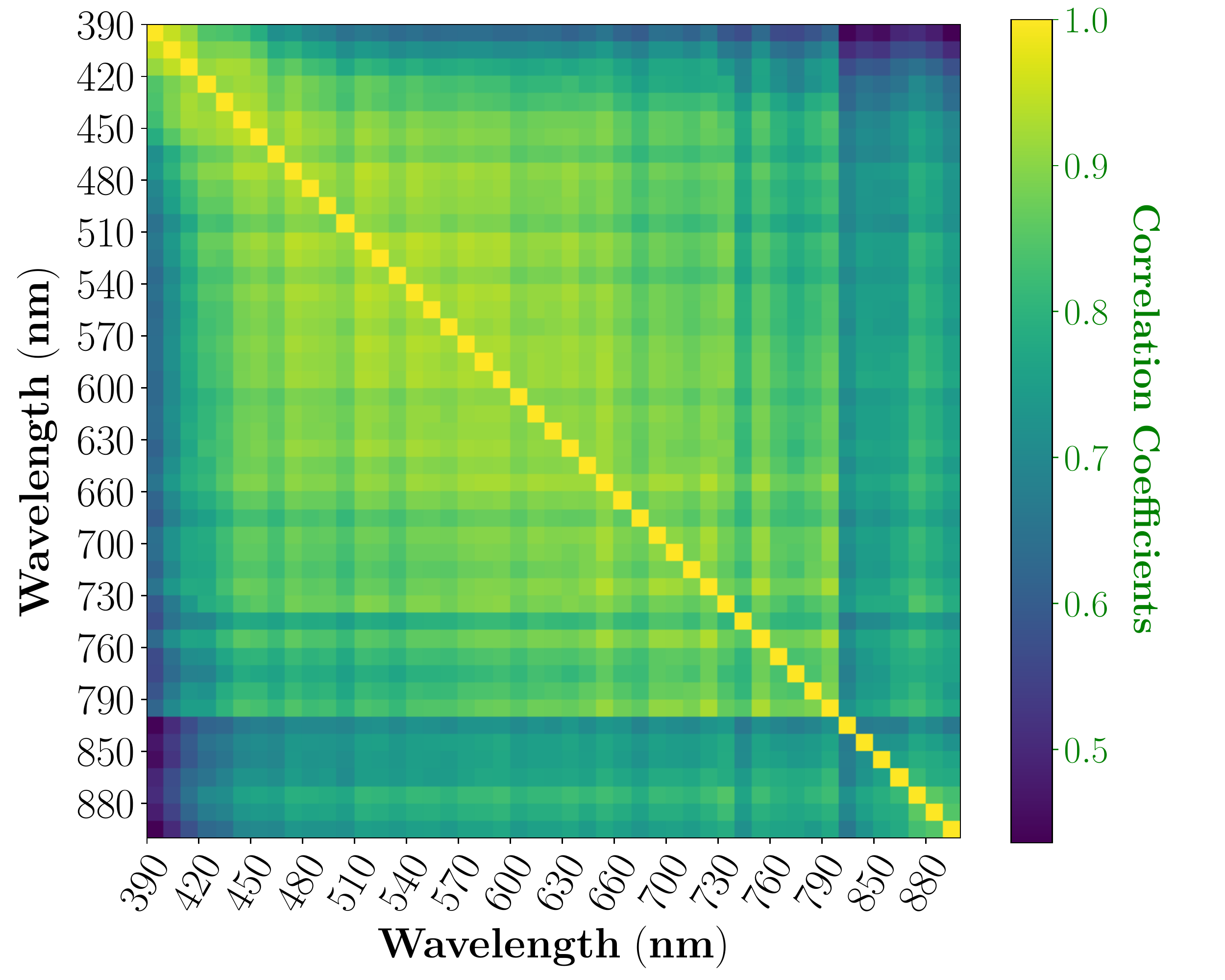}}
\end{minipage}
\begin{minipage}[t]{0.5\linewidth}
\centering
\resizebox{\hsize}{!}{\includegraphics{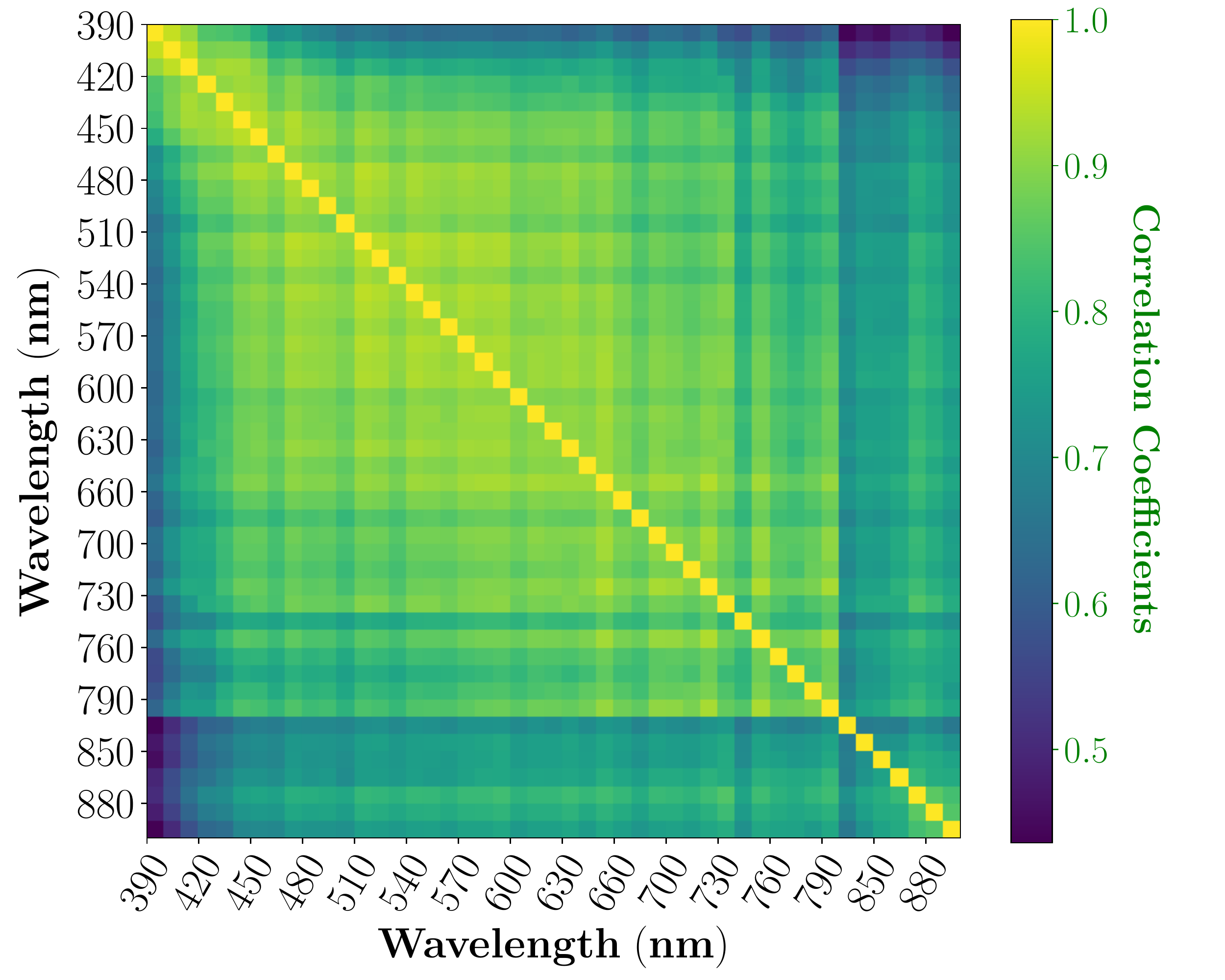}}
\end{minipage}
\caption{\small Correlation coefficients between small-scale flats of two filters.
The left and right panels are for the center and lower left corner regions, respectively.}
\label{Fig:pearson_r}
\end{figure*}

\begin{figure*}
\centering
    \includegraphics[width=15cm]{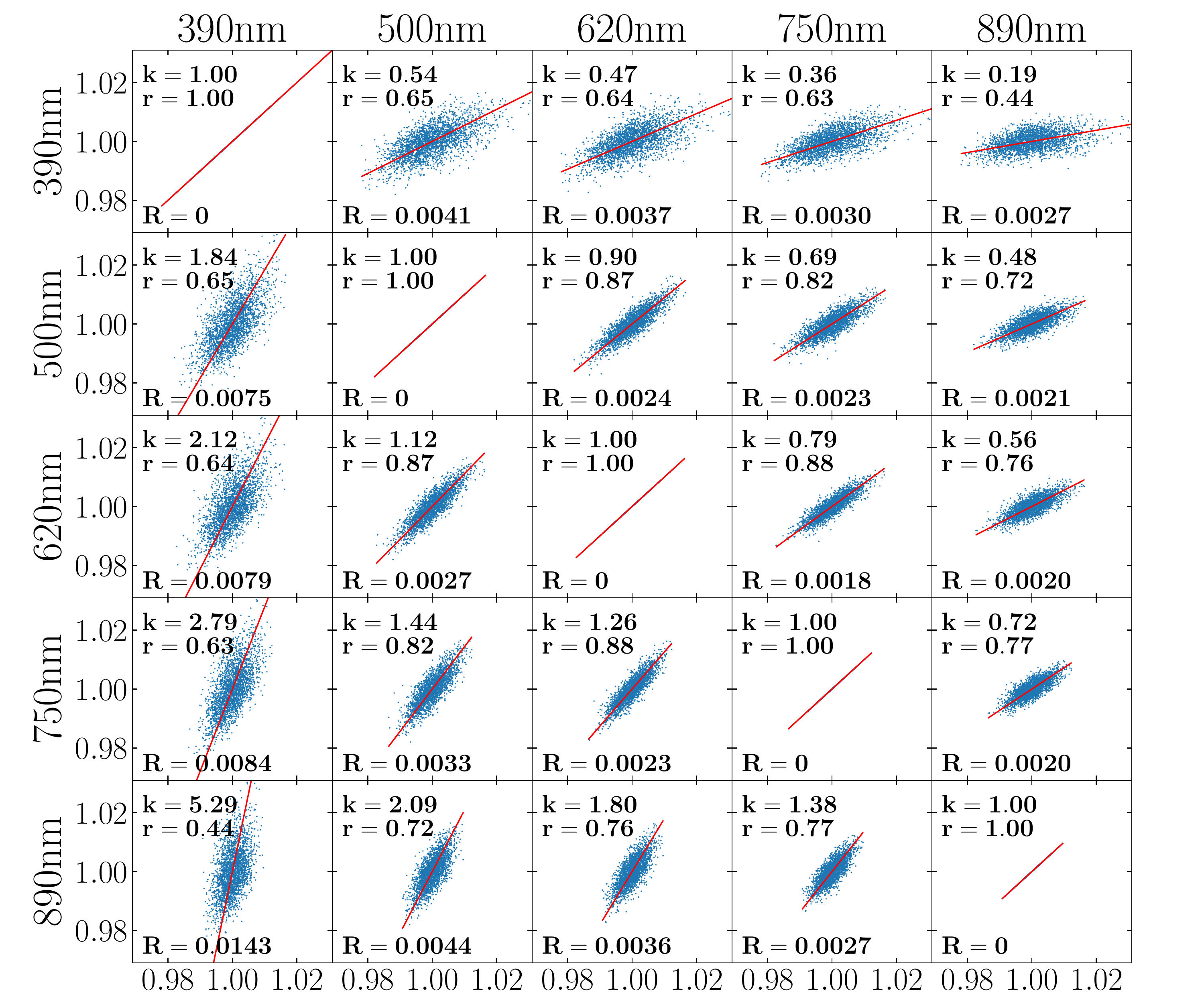}
    \caption{{\small Examples of correlation plots between small-scale flats of two filters in the center region. The filter central wavelengths are labeled. For each panel, the red line denotes the linear fitting result. The slope (k), correlation coefficient (r), and fitting residual (R) are marked.}}
    \label{Fig:fitting_5x5}
\end{figure*}

\begin{figure*}
\begin{minipage}[t]{0.5\linewidth}
\centering
\resizebox{\hsize}{!}{\includegraphics{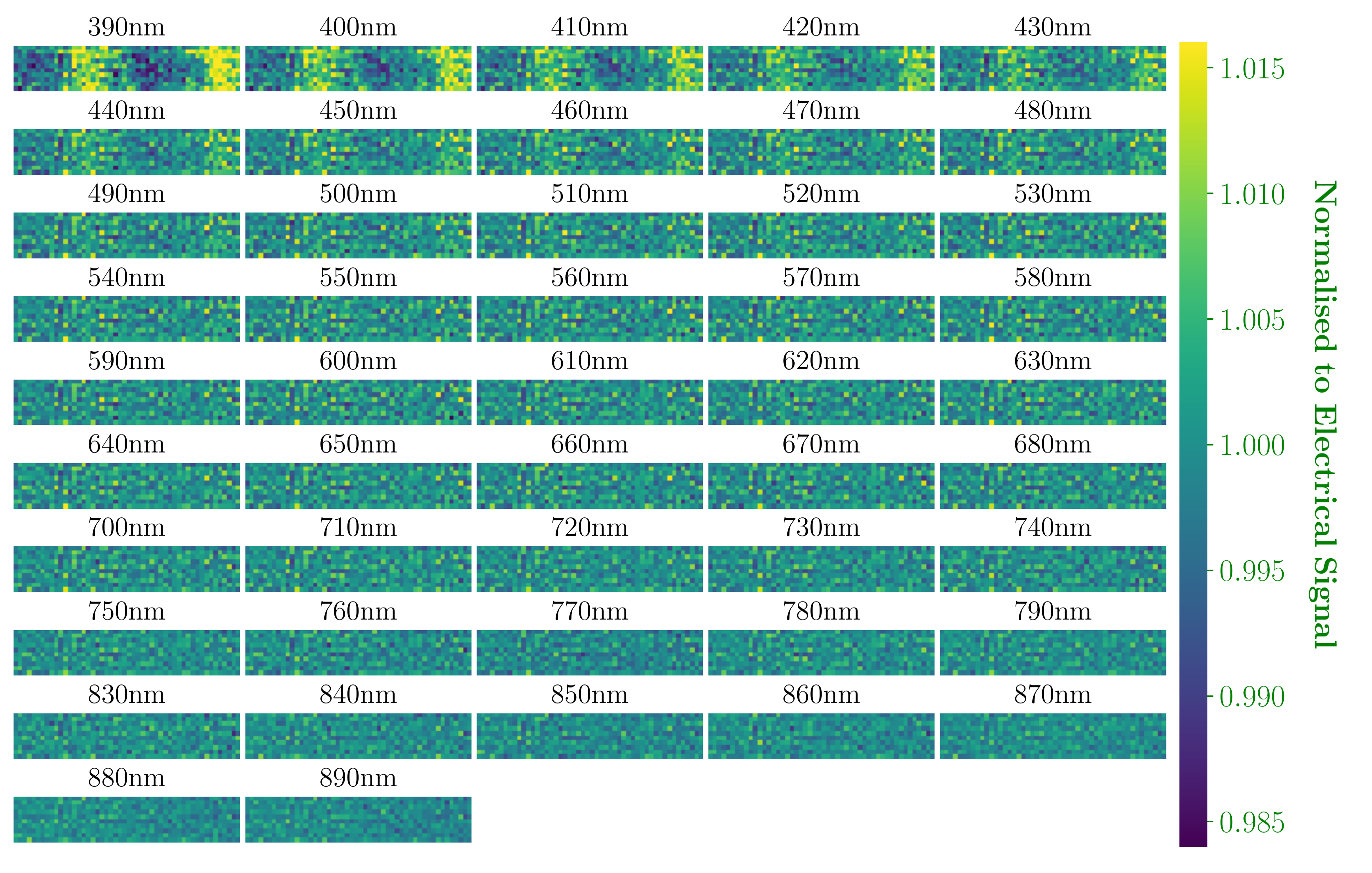}}
\end{minipage}
\begin{minipage}[t]{0.5\linewidth}
\centering
\resizebox{\hsize}{!}{\includegraphics{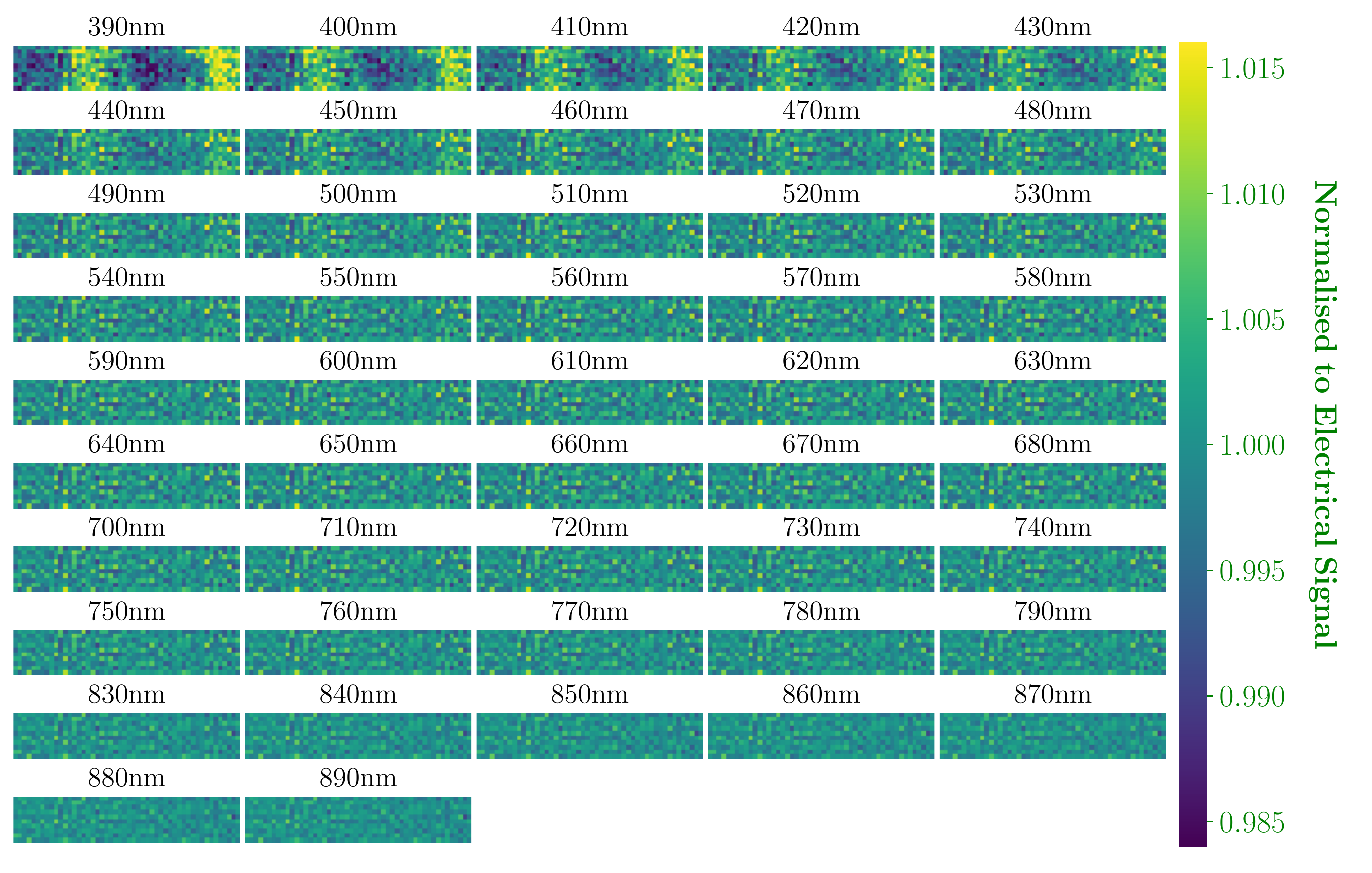}}
\end{minipage}
\caption{{\small Observed (left panel) and modelled (right panel) small-scale flats of the center region.}}
\label{Fig:obs_mod_500}
\end{figure*}

\begin{figure*}
\centering
\resizebox{\hsize}{!}{\includegraphics{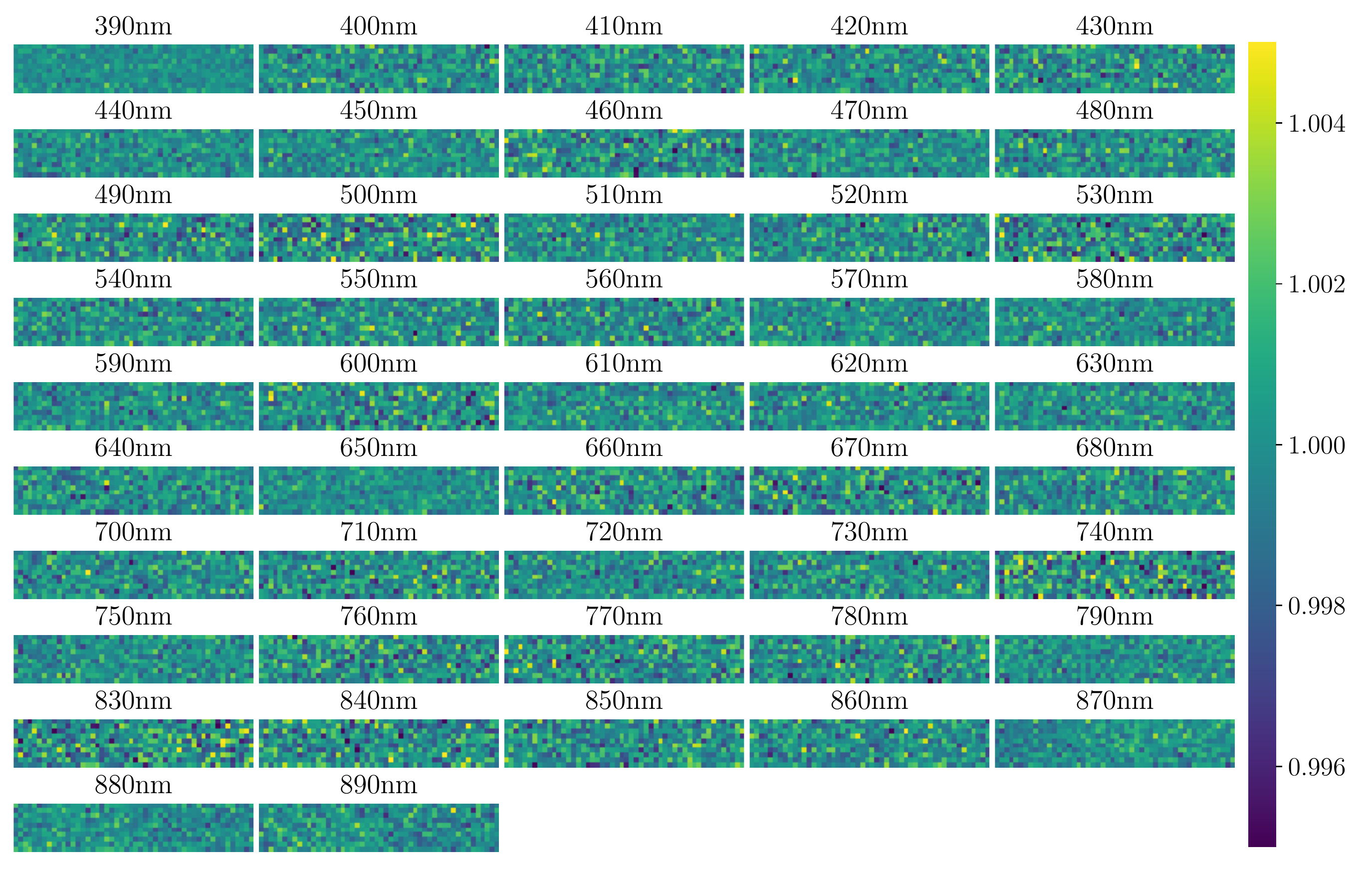}}
\caption{{\small Ratios of the observed to modelled small-scale flats of the center region.}}
\label{Fig:model_over_obs}
\end{figure*}

\begin{figure}
\centering
\resizebox{\hsize}{!}{\includegraphics{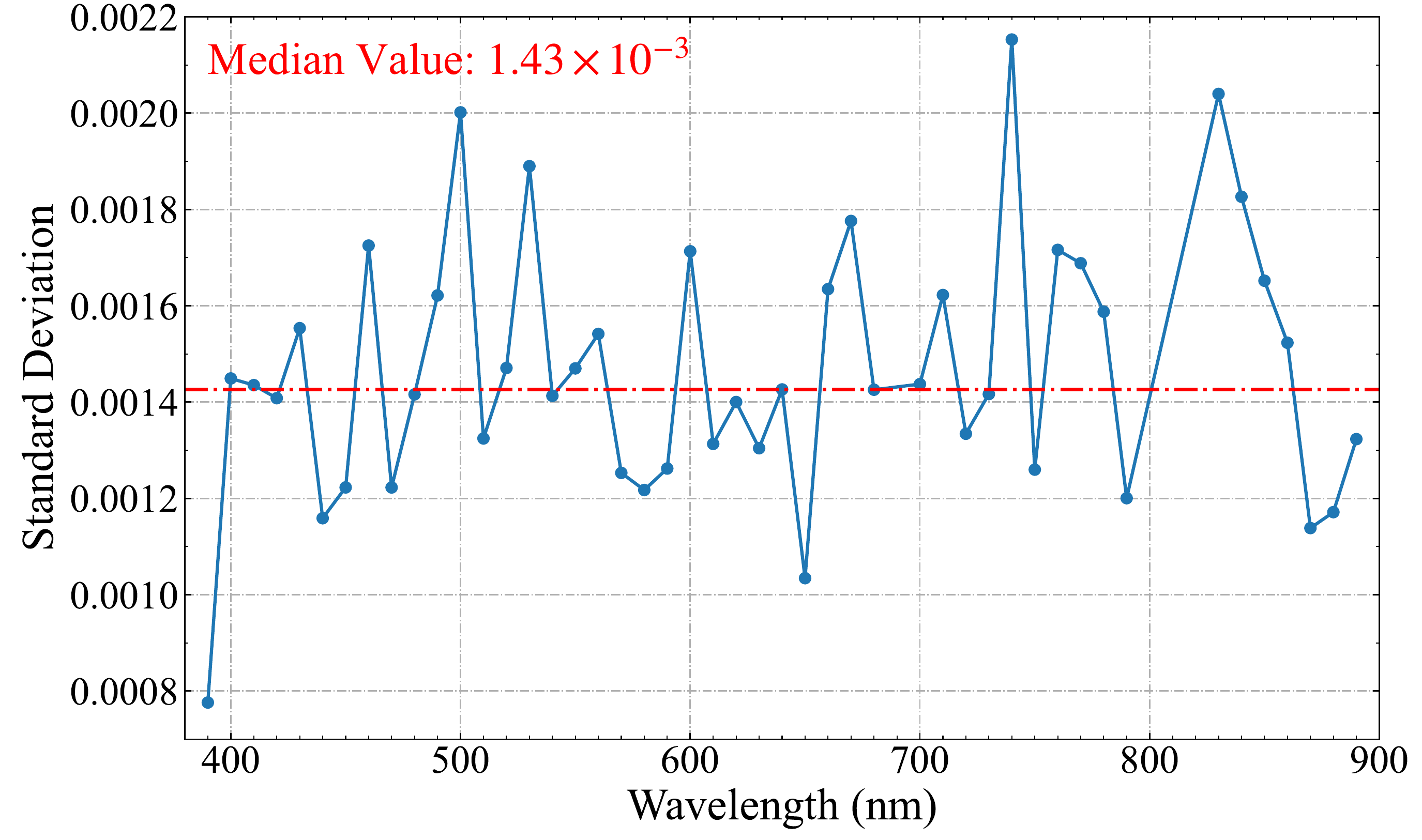}}
\caption{{\small Standard deviations of the ratios of the observed to modelled small-scale flats of the center region as a function of wavelength. The median value is also noted and shown as the red dotted line.}}
\label{Fig:sigma_model_over_obs}
\end{figure}

\begin{figure*}
\resizebox{\hsize}{!}{\includegraphics{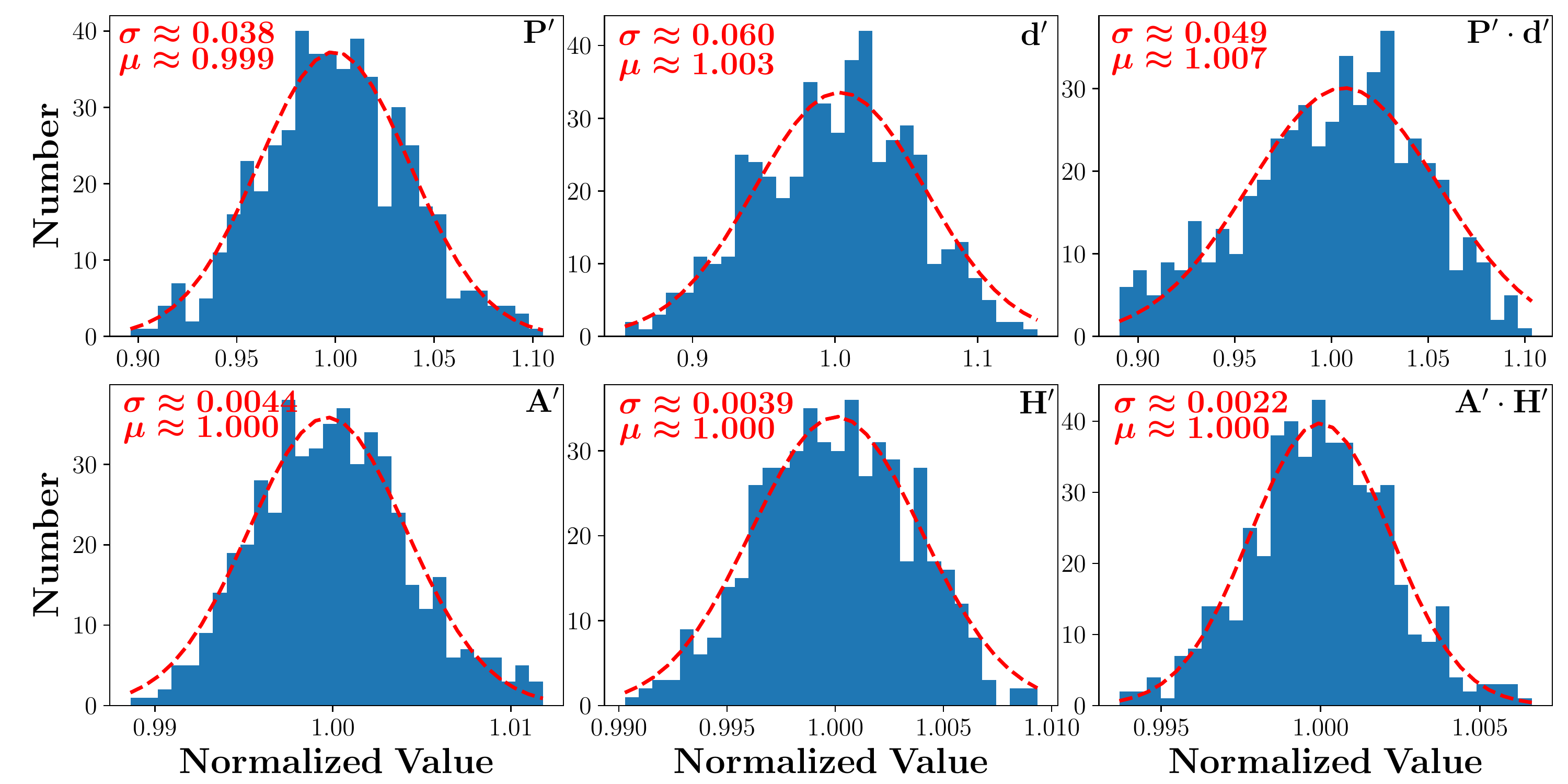}}
\caption{{\small Histogram distributions of different normalized model parameters, as labeled in the top right corner of each panel. The red dotted curves are Gaussian fitting results, with sigma values labeled in the top left corners.}}
\label{Fig:hist_paras}
\end{figure*}

\begin{figure}
\centering
\resizebox{\hsize}{!}{\includegraphics{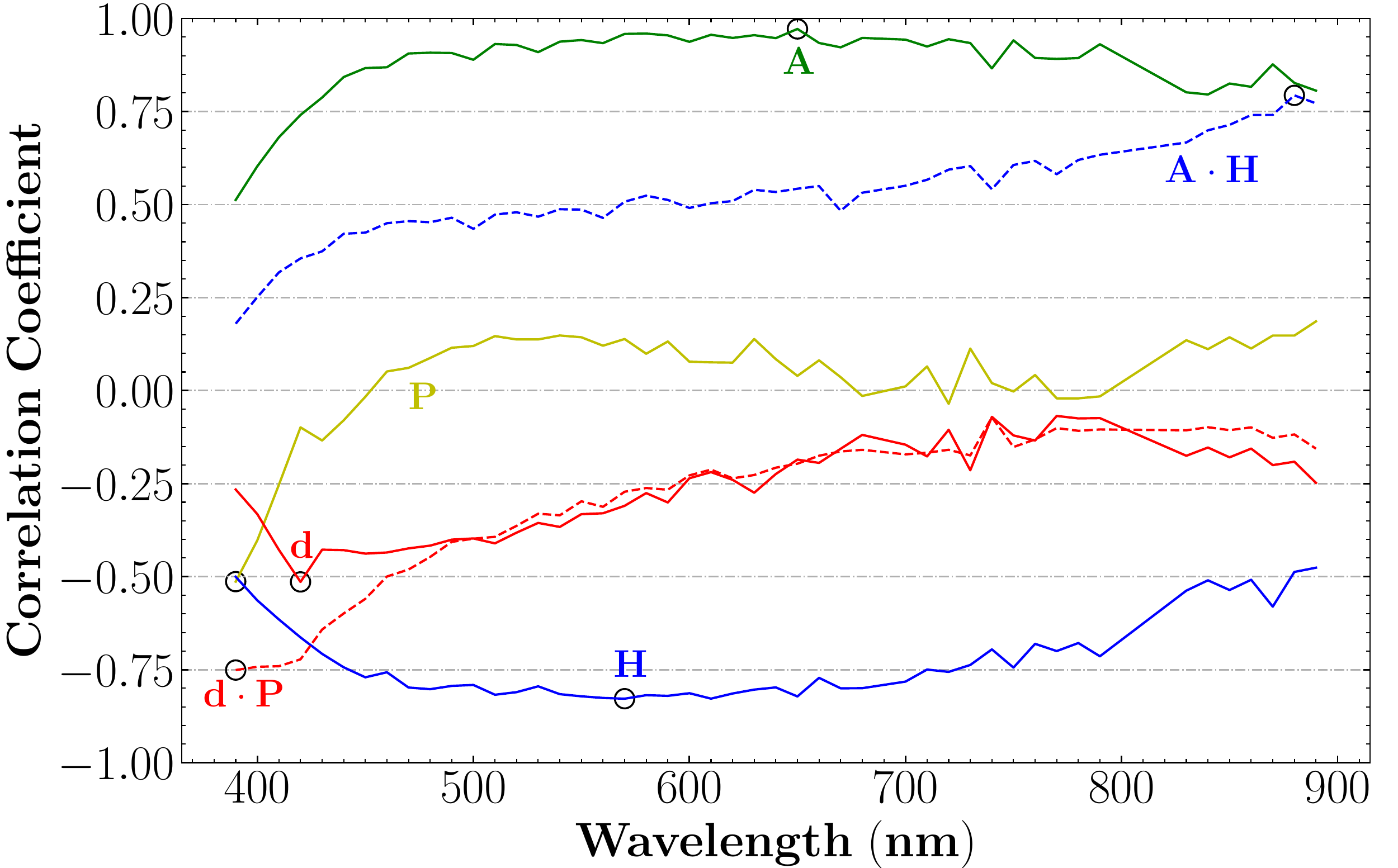}}
\caption{{\small Correlation coefficients between model parameters and observed small-scale flats as a function of wavelength.  }}
\label{Fig:pearson_r_paras}
\end{figure}

\begin{figure*}
\centering
\includegraphics[width=16cm]{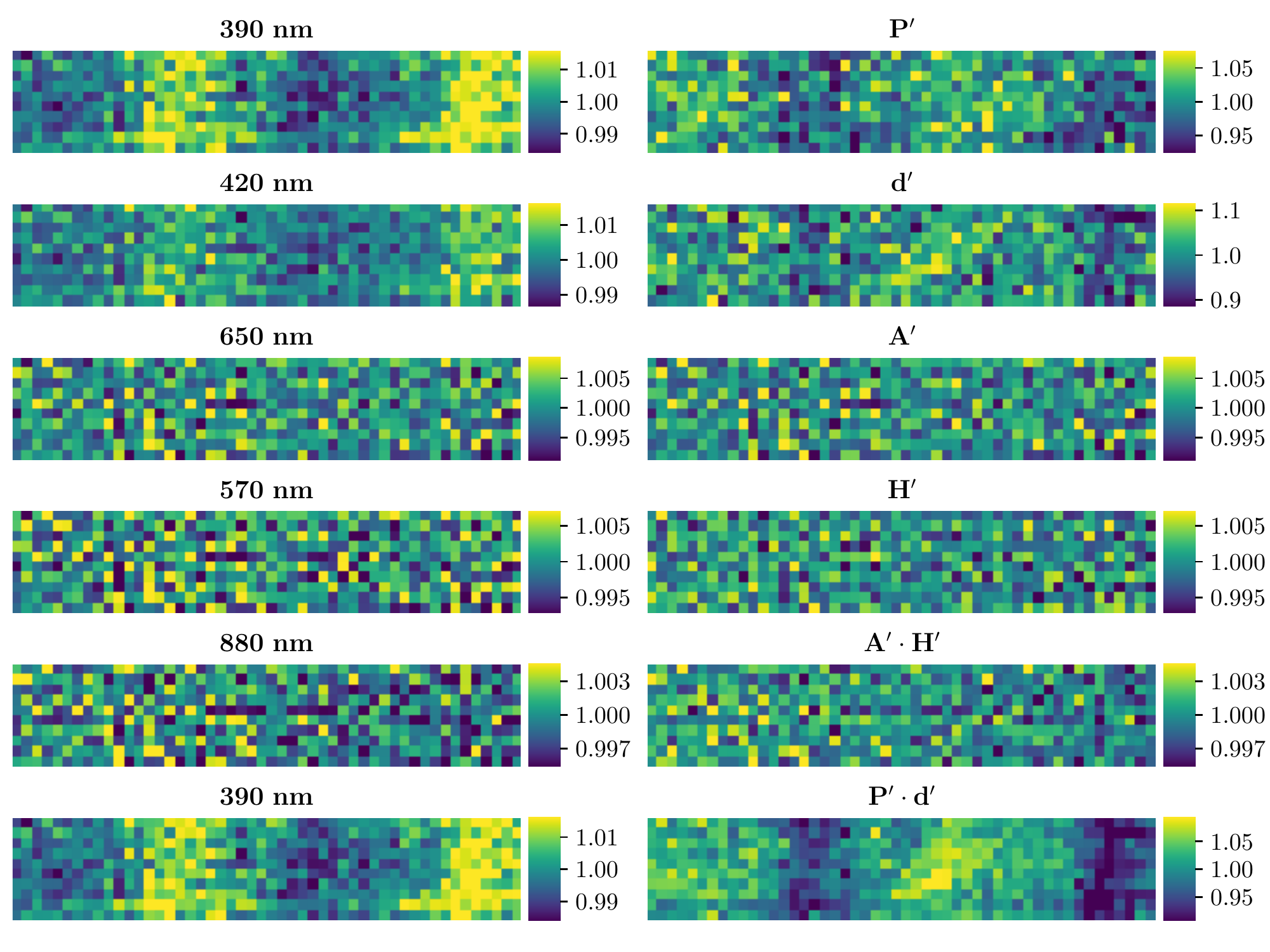}
\caption{{\small Model parameters (right panels) and their corresponding best correlated observed small-scale flats (left panels).}}
\label{Fig:paras}
\end{figure*}

\begin{figure*}
\centering
\includegraphics[width=14cm]{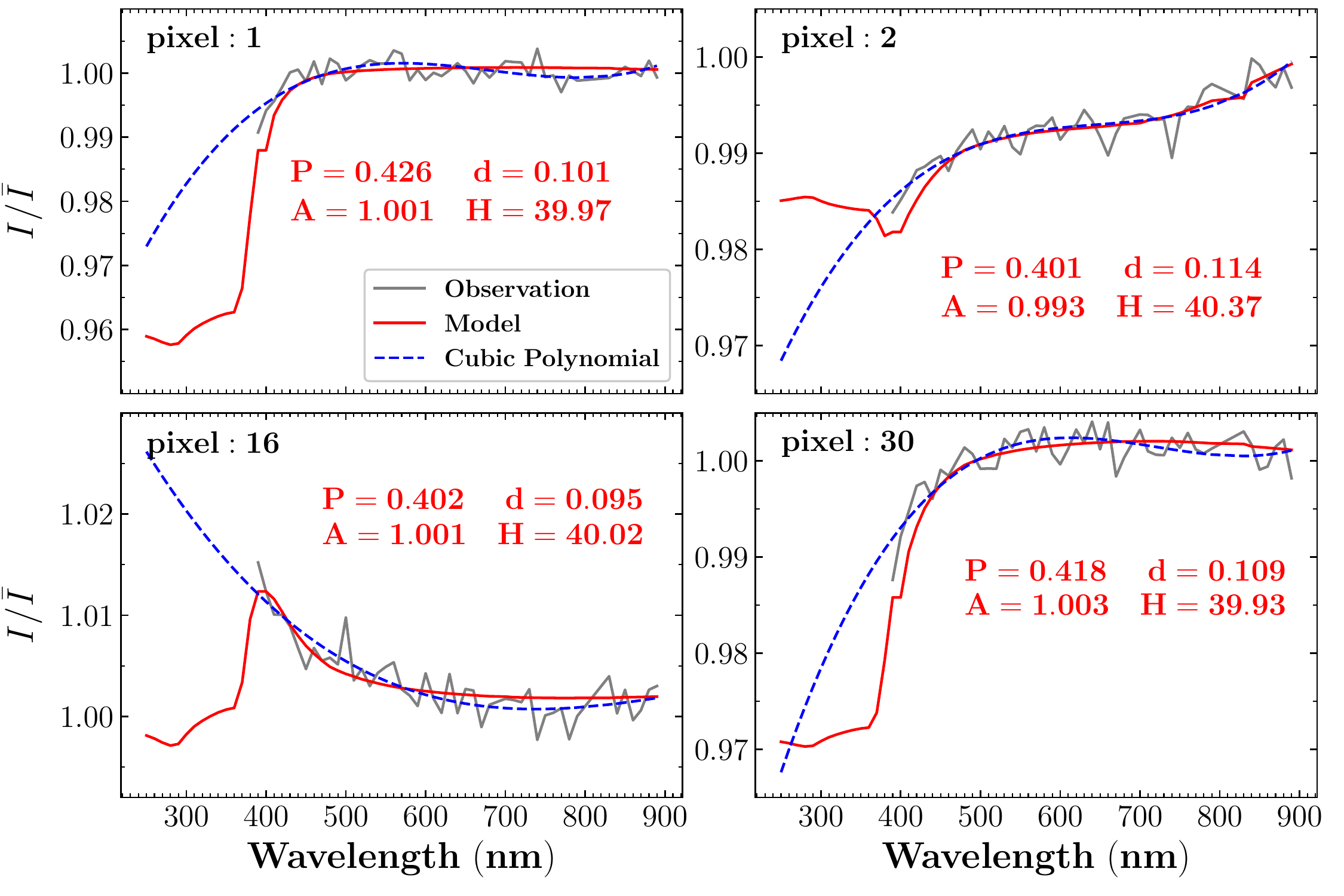}
\caption{{\small Observed (gray lines) and modelled (red lines) small-scale flats as a function of wavelength for four randomly selected pixels. The 4 model parameters for each pixel are labelled. For comparison, the results of cubic polynomial fitting are over-plotted in blue dotted lines. Note the significant discrepancies between the red and blue lines at shorter wavelengths.}}
\label{Fig:piexl_with_lambda}
\end{figure*}

\begin{figure*}
\resizebox{\hsize}{!}{\includegraphics{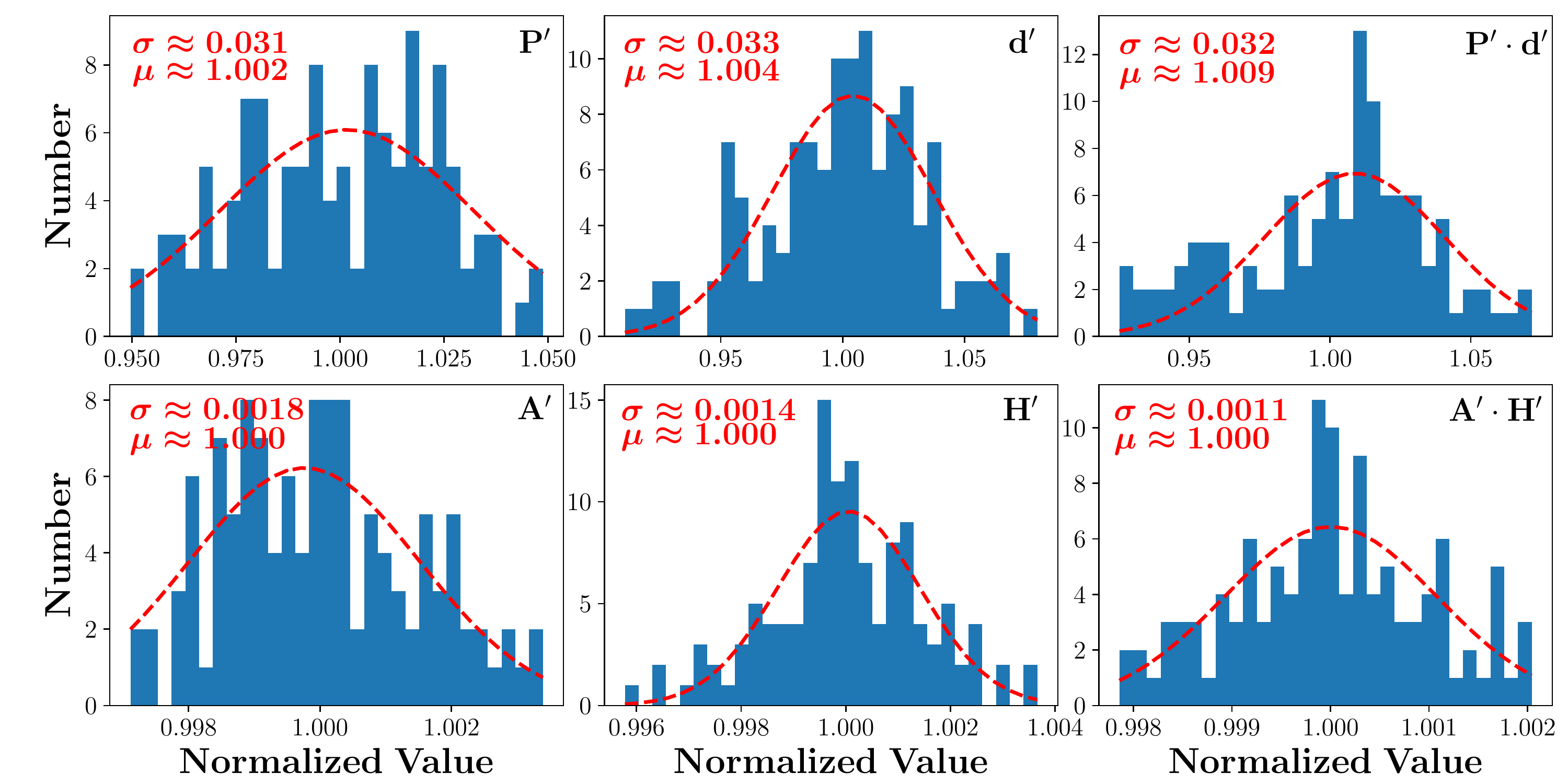}}
\caption{{\small  Histogram distributions of different normalized model parameters after $2\times2$ binning. 
The red dotted curves are Gaussian fitting results, with sigma values labeled in the top left corners.}}
\label{Fig:hist_paras_4×4}
\end{figure*}

\begin{figure}
\resizebox{\hsize}{!}{\includegraphics{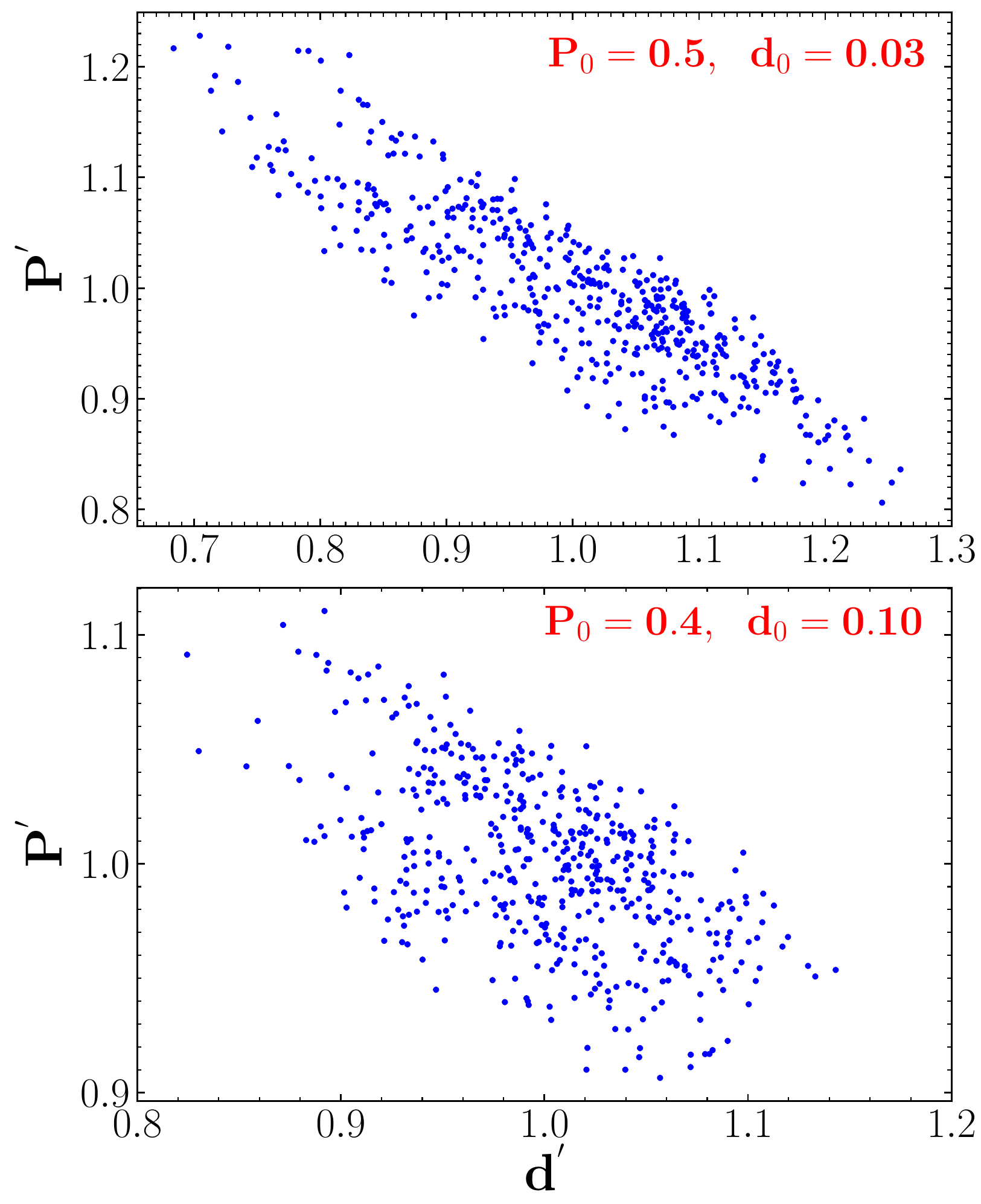}}
\caption{{\small Correlations between $P'$ and $d'$ for two sets of initial reference
values of $P$ and $d$. Top: ($P, d$) = (0.50, 0.03); Bottom: ($P, d$) = (0.40, 0.10).}}
\label{Fig:pearson_r_PH_and_dA}
\end{figure}

In order to determine the underlying parameters based on observational data, an initial set of
parameter values ($P,d,H,A$) are needed as reference, particularly $P$ and $d$.
$H$ and $A$ reference values are set to be 40\,$\mu m$ and 20\,$\mu m$\,$\times$\,20\,$\mu m$, respectively.
For $P$ and $d$, their reference values are constrained by the quantum efficiency curve
of the CCD used. This is because the quantum efficiency curve in blue wavelengths for a given CCD depends on not only
its reflectivity curve after coating, but also its typical $P$ and $d$ values. 
Figure\,\ref{Fig:QE_0401} plots the quantum efficiencies at four wavelengths of the mini-JPAS 
CCD. Predicted quantum efficiency curves of different [P,d] combinations are over-plotted,
assuming typical wavelength-dependent anti-reflectivity coating efficiency factors from Table 2.1 from \citet{2007ptd..book.....J}
\footnote{The true reflectivity curve is unknown.}. 
Finally, we choose the reference values of $P$ and $d$ to be 0.40 and 0.10 um, respectively. 
Due to the lack of measured quantum efficiencies and reflectivity curve, $P$ and $d$ reference 
values are not well constrained. We will discuss
the effect of different $P$ and $d$ reference values in Section \ref{sec:discussion}.

Given the reference values of the CCD, the $P,d,H,A$ values for each pixel are 
estimated by minimizing the differences between the observed and modelled small-scale flats.
A Python package for Sequential Least Squares Programming (SLSQP; \citealt{1988Tech...88...28}) is used in the optimization process. 

\section{Results} \label{sec:results}

Figure\,\ref{Fig:obs_2500} shows the small-scale flats we obtained at $47$ wavelengths. Two small regions of $50\times50$ pixels are selected, including one in the center and the other in the lower left corner of the CCD. For both regions, the bluer the filter, the larger the scatter. The trend is more clearly displayed in Figure\,\ref{Fig:sigma_flat}.
The standard deviation values decrease rapidly from about $1.0\%$ at 390\,$nm$ to $0.55\%$ at 440\,$nm$, then slowly to $0.3\%$ at 890\,$nm$. To investigate the effect of binning, the standard deviation values after another $2\times2$ binning are over-plotted. 
The values decrease significantly for wavelengths longer than around 450\,$nm$, but slightly for shorter wavelengths. 
This implies that if no binning was performed during the readout process, the standard deviation values would increase to some extent for wavelengths longer than around 450\,$nm$, but slightly for shorter wavelengths.
Note that the standard deviations for lower left corner are always slightly higher in 
the $4\times4$ binning. We have checked the standard deviations for the other three corners, it is not always the case that the values are higher in the corner regions.

Further, we perform a linear fitting for the normalized electrical signal sets at 2500 pixels between the small-scale flats from any two filters and estimated their correlation coefficients. The correlation coefficients are displayed in 
Figure\,\ref{Fig:pearson_r}. When the two filters are closer in wavelength, the correlations are usually stronger,  and the slopes are closer to $1$. Note that the grid pattern in Figure\,\ref{Fig:pearson_r} is not real. It is mainly due to the relatively lower SNRs (numbers of exposure times) of flats of several filters (e.g., at 460\,$nm$, 530\,$nm$, 600\,$nm$, 670\,$nm$, 740\,$nm$, 830\,$nm$, 840\,$nm$ and 850\,$nm$). Examples of the correlations are shown in Figure\,\ref{Fig:fitting_5x5}. 
The result suggests that the small-scale flats are more similar when the center wavelengths are closer.

We now apply our model to fit the observed small-scale flats.
In order to reduce the number of free parameters and computing time, 
we further select a smaller region of 10$\times$50 pixels 
from the left panel of Figure\,\ref{Fig:obs_2500}, as shown in the left panel of Figure\,\ref{Fig:obs_mod_500}\footnote{We have tested several different regions, and the results are similar.}.
The selected data is used to constrain our model, which has a total of 10$\times$50$\times$4 free parameters.  
The right panel of Figure\,\ref{Fig:obs_mod_500} shows the model results, which agree well 
with observations.
Detailed comparisons between the observed and modelled 
flats are displayed in Figure\,\ref{Fig:model_over_obs}. The standard deviations of their ratios are plotted in Figure\,\ref{Fig:sigma_model_over_obs}. It can be seen that our model reproduces the observed small-scale flats well 
in all the wavelengths. The median standard deviation is only $0.14\%$.
Note that the standard deviations show a moderate anti-correlation with the numbers of exposure times, suggesting that a large fraction of the scatters are contributed by random errors in the flats. 

The model parameters are normalized by their corresponding reference values.
Figure\,\ref{Fig:hist_paras} shows histogram distributions of the four normalized parameters ($P'$, $d'$, $A'$, and $H'$). The normalized parameters follow Gaussian distributions roughly, with sigma values of 3.8\%, 6.0\%, 0.44\% and 0.39\% for $P'$, $d'$, $A'$, and $H'$, respectively. The distributions of $P'\cdot d'$ and $A'\cdot H'$ 
are also plotted in Figure\,\ref{Fig:hist_paras}, with sigma values of  4.9\% and 0.22\%, respectively.

To investigate sensitivities of different model parameters on flats in different wavelengths, 
the correlation coefficients between model parameters ($P$, $d$, $A$, $H$, $P\cdot d$, and $A\cdot H$) 
and observed small-scale flats are plotted against wavelength in 
Figure\,\ref{Fig:pearson_r_paras}. 
The correlation coefficient for the $A\cdot H$ parameter increases with wavelength, reaching a peak value 
of about 0.75 at 880\,$nm$. 
The correlation coefficient for the area parameter $A$ peaks at a medium wavelength around 650\,$nm$. 
The correlation coefficients for both $P$ and $d$ parameters are small for all the available wavelengths.
While the $P\cdot d$ parameter has a strong negative correlation with the flat at 390\,$nm$. To better demonstrate the correlations, the spatial distributions of model parameters and their best correlated 
flats are compared in Figure\,\ref{Fig:paras}. 
The correlation results are as expected:
\begin{enumerate}
  \item At wavelengths where the photon absorption length $L$ is smaller than or comparable to the electron absorption depth $d$ (shorter than about 390\,$nm$ in the case of this work where $d=0.10$\,$\mu m$) the small-scale flat mainly comes from the effect of charge collection efficiency, depending on both
   the electron absorption probability $P$ and $d$. 
   \item At wavelengths where $L >> d$ and $L << H$, around 650\,$nm$ in the case of this work, the effect of charge collection efficiency is small and the CCD pixel thickness $H$ does not matter, the small-scale flat is dominated by variations of the pixel effective area $A$.
  \item At wavelengths where $L$ is comparable to or larger than $H$, the effect of charge collection efficiency is small and the CCD pixel thickness starts to matter, the small-scale flat mainly comes from variations of $H\cdot A$. 
 \end{enumerate}
    
In this work, we also performed a cubic polynomial fitting to wavelength-dependent small-scale flats for each pixel and compared the results with those of the physical model. Note that both methods have 4 free parameters for each pixel. 
The results are compared in Figure\,\ref{Fig:piexl_with_lambda} for 4 randomly selected pixels. 
Both methods work well for the observed wavelength range. However, for shorter wavelengths ($<$ 400\,$nm$), the results 
differs significantly. If our physical model is correct, it suggests that extrapolation of polynomial fitting results to shorter wavelengths are unreliable due to rapid variations of the charge collection efficiencies.

\section{Discussion} \label{sec:discussion}
The mini-JPAS camera does show wavelength-dependent small-scale flats, which 
are well explained by our model.
The small-scale flats of 
the mini-JPAS camera come from at least two aspects. One is inhomogeneities 
of the quantum efficiency (particularly charge collection efficiency) between adjacent CCD pixels. 
The other is variations of the effective area and thickness of CCD pixels. The former
dominates small-scale flats in short wavelengths while the latter in longer (visual and infrared) wavelengths. 

The relative variation of effective area between different pixels has a typical value of
$0.44\%$ (Figure\,\ref{Fig:hist_paras}), which cannot be ignored when precise photometric, 
astrometric and shape measurements of astronomical objects are needed.
Considering that the pixel scale is 0.46 arcsec per pixel (0.23 arcsec per physical pixel) for the mini-JPAS camera
and a typical seeing of 0.71 arcsec for the OAJ site (\citealt{2010PASP..122..363M}), the effect of
pixel area variations on photometry due to inappropriate small-scale flat-fielding is probably small (at a few mmag level).  
We note a strong anti-correlation between $A'$ and $H'$ parameters.
As shown in Figure\,\ref{Fig:hist_paras}, the  $A'\cdot H'$ has a much smaller dispersion value
than those of $A'$ and $H'$ independently. The result suggests that volumes of CCD pixels are more uniform than their areas and thicknesses.   

Note that a pixel in this work corresponds to four physical pixels due to the binning in the readout process. 
To investigate the effect of binning on this work, we plot histogram distributions of different normalized model parameters 
after $2\times2$ binning in Figure\,\ref{Fig:hist_paras_4×4}. 
For parameters $A'$, $H'$ and  $A'\cdot H'$, the sigma values decrease significantly by about a factor of two. 
It is not surprising, as one would expect that variations of pixel area between neighbor pixels 
are anti-correlated (\citealt{2017PASP..129h4502B}). Such anti-correlations can also be seen in the right panels of Figure\,\ref{Fig:paras}. 
Therefore, we can infer that the true variations of $A'$, $H'$ and  $A'\cdot H'$ parameters for individual physical pixels 
are likely much larger than those shown in Figure\,\ref{Fig:hist_paras}.
For $P'$, the sigma value decreases slightly from 3.8\% to 3.1\%, suggesting a much weaker binning effect. 
It is probably because $P$ has a larger variation scale, as can be seen in the top panels of Figure\,\ref{Fig:paras}. 

Due to the dependence of small-scale flats on wavelength, flat-fielding in the traditional way, which 
depends only on filters, may cause color terms in photometric calibration, particularly in blue and ultraviolet filters.
Small-scale flat-fielding in slit-less spectroscopic surveys may suffer similar problems. 
In order to achieve high-precision photometric/flux calibration, detailed modelling of 
wavelength-dependent variations of small-scale flats is needed. 
In this case, obtaining well measured flats in a number of narrow-band filters are necessary.
The central wavelengths of selected narrow-band filters slightly change with the quantum efficiency curve of CCD. A good sampling in the blue and ultraviolet wavelengths, 
where quantum efficiency varies rapidly, is suggested.

We adopted a set of reference values $P,d$=[0.40, 0.10] in this work. 
To test the effect of different reference values,we selected another set of values $P,d$=[0.50, 0.03] and compared them with the current results. 
Figure\,\ref{Fig:pearson_r_PH_and_dA} plots correlations 
between $P$ and $d$ parameters for the two sets of reference values.
A strong correlation between $P$ and $d$ is seen for $P,d$=[0.50, 0.03] in the bottom panel, 
while the correlation is much weaker for $P,d$=[0.40, 0.10] in the top panel. 
The reason is that flats of short wavelengths, whose photon absorption depths are smaller than 
$d$, are needed to break the degeneracy between $P$ and $d$. $d$ is larger in the top panel (0.10\,$\mu m$) than that in the bottom (0.03\,$\mu m$), therefore the correlation is weaker. The result demonstrates the importance of flats at very short wavelengths.

\section{Conclusions} \label{sec:conclusions}
Using the unique data set provided by the flat-fields of 47 narrow-band filters
used by the JPAS-{\it Pathfinder}, 
this paper addressed the question: how do the small-scale flats of a CCD camera depend on 
wavelength? Observationally, we detect variations from small-scale flats from different filters.
The variations are stronger in shorter wavelengths.
Small-scale flats of two filters close in central wavelengths are correlated, and we find that the closer the wavelengths, the stronger the correlation. Theoretically, we use a simple physical model 
to explain the observed wavelength-dependent variations of small-scale flats. 
The model considers the variations of 
charge collection efficiencies, effective areas and thicknesses between pixels, 
with four free parameters ($P,d,H,A$) to characterize each pixel. 
The observations are successfully reproduced to a precision of about $0.14\%$.
    
The model result shows that the wavelength-dependent variations of small-scale flats of 
the mini-JPAS camera originate from two aspects. On one hand, the inhomogeneities 
of the quantum efficiency (particularly charge collection efficiency) between different CCD pixels dominate the variations at short wavelengths.
On the other hand, the variations of the effective area and thickness of CCD pixels, are more 
important in longer (visual and infrared) wavelengths. 
The relative variation of effective area between different pixels has a typical value of
$0.44\%$, which cannot be ignored during flat-fielding  
when high-precision photometric, astrometric, or shape measurements of astronomical objects are needed.
In order to achieve high-precision photometric calibration for imaging surveys, or flux calibration for 
slit-less spectroscopic surveys, detailed modelling of wavelength-dependent variations of small-scale flats 
is also needed to avoid color dependent corrections, particularly in short wavelengths 
where CCD quantum efficiency curve varies rapidly. 

In order to model the wavelength-dependent variations of small-scale flats, we find that different parameters
are sensitive to flats of different wavelengths. 
At wavelengths where the photon absorption length $L$ is smaller than or comparable to the electron absorption depth $d$,  the small-scale flat mainly comes from the effect of charge collection efficiency, depending on both
the electron absorption probability $P$ and $d$. 
At wavelengths where $L >> d$ and $L << H$, the small-scale flat is dominated by variations of the pixel 
effective area $A$. At wavelengths where $L$ is comparable to or larger than $H$, the small-scale flat mainly comes from variations of $H\cdot A$. Therefore, a small number (around ten) of small-scale flats with well-selected wavelengths are sufficient to reconstruct small-scale flats in other wavelengths.

\begin{acknowledgments}
We acknowledge the anonymous referee for his/her valuable comments that improve the quality of this paper.
The PRNU model adopted in this work was developed as part of Mr. Baocun Chen's undergraduate thesis work under the supervision of H.~Zhan. 
We acknowledge Drs. Stavros Akras, Alvaro Alvarez-Candal, Luis Alberto D{\'i}az Garc{\'i}a, and Zhenya Zheng for a careful reading of the manuscript. 
This work is supported by the National Natural Science Foundation of China through the project NSFC 11603002,
the National Key Basic R\&D Program of China via 2019YFA0405503 and Beijing Normal University grant No. 310232102. 
We acknowledge the science research grants from the China Manned Space Project with NO. CMS-CSST-2021-A08 and CMS-CSST-2021-A09.
J. V. acknowledges the technical members of the UPAD for their invaluable work: Juan Castillo, Tamara Civera, Javier Hernández, Ángel López, Alberto Moreno, and David Muniesa.

Based on observations made with the JST/T250 telescope and JPCam at the Observatorio Astrofísico de Javalambre (OAJ), in Teruel, owned, managed, and operated by the Centro de Estudios de F{\'i}sica del Cosmos de Arag{\'o}n (CEFCA). We acknowledge the OAJ Data Processing and Archiving Unit (UPAD) for reducing and calibrating the OAJ data used in this work.

Funding for the J-PAS Project has been provided by the Governments of Spain and Aragón through the Fondo de Inversi{\'o}n de Teruel, European FEDER funding and the Spanish Ministry of Science, Innovation and Universities, and by the Brazilian agencies FINEP, FAPESP, FAPERJ and by the National Observatory of Brazil. Additional funding was also provided by the Tartu Observatory and by the J-PAS Chinese Astronomical Consortium.

\end{acknowledgments}

\end{document}